\documentclass[prodmode]{acmsmall}

\usepackage{amsmath}

\usepackage{natbib}
\usepackage{graphicx}

\usepackage{amsmath}
\usepackage{amssymb}

\usepackage{times}

\usepackage{pifont}
\usepackage{epsfig}
\usepackage{tabularx}
\usepackage{amssymb}
\usepackage{latexsym}
\usepackage{amsmath}
\usepackage{delarray}
\usepackage{moreverb}
\usepackage{xspace}

\usepackage{lscape}
\usepackage{changebar}
\usepackage{graphicx}
\usepackage{graphics}
\usepackage{mflogo}
\usepackage{xspace}
\usepackage{texnames}
\usepackage{rotating}
\usepackage{alltt}

\usepackage{amsmath}
\usepackage{epsfig}
\usepackage{booktabs,paralist}

\newcommand{\pred}{{}}

\newcommand{\f}{\varphi}

\newcommand{\X}{\mathbb{X}}

\newcommand{\R}{\mathbb{R}}

\newcommand{\F}{\mathbb{F}}

\newcommand{\Vc}[1]{{\boldsymbol #1}}

\newcommand{\mypar}[1]{{\bf #1.}}

\newcommand{\doublefigure}[5]{{\begin{figure}%
\centering %
\includegraphics[width=#1\linewidth]{#2}
\includegraphics[width=#1\linewidth]{#3}
\caption{#4}
\label{#5}%
\end{figure}}}

\newcommand{\Doublefigure}[5]{{\begin{figure} %
\centering %
\includegraphics[width=#1\linewidth]{#2}%
\includegraphics[width=#1\linewidth]{#3}%
\caption{#4}%
\label{#5}%
\end{figure}}}

\newcommand{\singlefigure}[4]{{\begin{figure}[htb] %
\centering %
\includegraphics[width=#1\linewidth]{#2}%
\caption{#3}%
\label{#4}%
\end{figure}}}

\newcommand{\orthogonal}{%
\mathrel{\raisebox{.1em}{%
\reflectbox{\rotatebox[origin=c]{90}{$\models$}}}}}

\begin{document}

\markboth{D'Alberto, Milenkiy, and Azizi}{Mobile Visit Lifts}


\title{Digital Advertising: the Measure of Mobile Visits Lifts }

\author{PAOLO D'ALBERTO
  \affil{...} 
  VERONICA MILENKIY
  \affil{...}
  FAIRIZ FI AZIZI
  \affil{...}
}

\author{ Paolo D'Alberto, Veronica Milenkiy,  \and Fairiz Fi Azizi}

\begin{abstract}
Mobile-phone advertising enables marketers to reach customers at a
personal level and it enables the measure of costumers' reaction by
novel approaches, in real time, and at scale. By keeping a device
anonymous, we can deliver custom adverts and we can check when the
device owner will visit a specific mortar-and-brick location.  This is
the first step in a sale. By measuring visits and sales, the original
marketers can determine their return on advertising and they can prove
the efficacy of the marketing investments.  We turn our attention to
the measure of lift: we define it as the visit acceleration during the
campaign flight with respect to a controlled baseline. We present a
theoretical description; we describe a general and a simplified
approach in composing the exposed and the control baseline; we develop
two different vertical approaches with different comparable solutions;
finally, we present how to carry the experiments and the measures for
a few dozens campaigns; these campaigns range from hundred thousands
devices and counting a few hundred visits to a handful locations, to
sixty million devices and counting million visits to thousands
locations. We care about experiments at scale.
\end{abstract}




\acmformat{P. D'Alberto, Veronica Milenkiy, and Fariz Fi Azizi. Visit
  Lifts}

\begin{bottomstuff}
This paper has about 8 patents granted.  
\end{bottomstuff}

\maketitle


\section{Introduction}
\label{sec:introduction}

Advertising reaches customers with propositions and suggestions to
appeal the features of a product to a tailored clientele in order to
increase the product acceptance and its craftsmen revenues. If the
customer has a mobile device and has been exposed to any adverts, we
can measure their influence by counting any form of active actions
such as visiting brick-and-mortar locations. In practice, advertising
is a social experiment at large scale.  Differently from historical social
experiments or medical trials, we do not have scale problems and we
have a rich often-continuous feature space describing our exposed and
control groups. As in a social experiment, we often have a clear intent for
the experiment but we may have limited or poor means to measures the
effect of the experiment. Common questions are: did the advertising
campaign work?  How much did it work? How can we measure visits,
goals? How can we claim that the experiment brought more visits than a
control baseline?  We shall address most of these questions in a
constructive way: First, we shall propose one measure and two
methodologies; second, we shall describe how the methodologies are
related (one is more general than the other); and third, we present
experiment results and quantitative measures for the two different
approaches. In the following paragraph, we shall sketch an intuitive
outline.

We measure a campaign goals by the {\bf lift}, which is a visit
acceleration. We use a mental exercise as introduction. Two authors
Paolo and Fi are reached by a campaign advert on the same day (e.g.,
January 21, 2017), to play golf at Pebble Beach during the next month
of February.  Fi plays regularly (at pebble) and Paolo is no good. Fi
is an example of re-targeting customers; that is, we expose a customer
already interested.  Paolo is at best a new comer.  Veronica was not
exposed but she regularly interacts with both Paolo and Fi: she is a
candidate for control because she could have being exposed on the same
day (January 21, 2017) but she was not then and thereafter.
Intuitively, the campaign has an effect if Fi will go more often than
before or Paolo will tee once or more. Inherently, there is a concept
of acceleration; that is, exposed has to do more after exposure and
more than who is not exposed.  Veronica will give a reference how hard
was to achieve the above goal.  If Veronica plays regularly, Fi should
also be compared to Veronica. If they appear teeing off at the same
pace then they may prepare for the masters' (more) or avoiding the
green because an incoming blizzard (less).  If Veronica does not play
but Paolo's excitement makes her aware of the beauty of the game,
Paolo should be compared to Veronica.  Our measure of lift must
capture this desire to change the pace of visits as time unfold, this
is why we use the term of acceleration. Of course, we must consider an
average acceleration and do not forget this is a an intuitive but
gross oversimplification.

The previous mental exercise introduces acceleration in combination
with matching across comparable people. How can we describe customers
to draw a comparison? In the example above, if we know that Fi has
visited before exposure and Veronica did as well, Fi and Veronica are
{\em better} comparable than Paolo and Veronica. The visits before
exposure is a discriminating features. If the campaign targets only
males, Veronica would not be targeted and, thus, we should not use her
as control baseline.  Some targeting features are obtained by
voluntary identification and others are the results of
approximations. We represent these approximated features using
continuous probabilities. In the best scenarios, we are targeting
many-to-many matching and thus once again aiming to the computation of
{\em average} accelerations.

For third party campaigns, we do not know anything about the campaign
targeting and we must infer it by the people targeted. In the example,
Veronica is the closest control baseline to Fi or to Paolo as a
function of the features used: previous visits or gender. Interesting,
there are different approaches to matching and different matching
algorithms will provide different results. In this work, we encourage
the application of different methodologies and matching algorithms.

The organization of the paper follows.  In Section
\ref{sec:definition}, we introduce our notations and the definition of
    {\em Lift}. We present also two interpretations: a balanced
    approach in Section \ref{sec:CMS} and an unbalanced one in Section
    \ref{sec:control-selection}; we show results for both. We present
    our original contribution for the features computation in a
    continuous space in Section \ref{sec:keywords}, which is composed
    of a {\em Location Graph}, Section \ref{sec:location-graph}, and a
    {\em User Profile}, Section \ref{sec:user-profile}. We present our
    original contribution about {\em Matching} in Section
    \ref{sec:matching-intro}. We present experimental results in
    Section \ref{sec:experiments} and we conclude in Section
    \ref{sec:conclusions}.

\section{Problem Definition and Introductory Notations}
\label{sec:definition}

We start by presenting yet another example: we start an advertising
campaign where we invite the {\bf exposed people} to visit a coffee
shop. The exposure is by means of digital advertising and the {\bf
  visit} is by means of a distance measure between the mobile devices
to coffee shops. The goal of any advertising is to affect visits so
that the exposed group has more visits than a {\bf control group} and,
because exposed and control can be quite different in size, a better
visit rate. We may have very different ways to choose the
exposed--control set. We distinguish two ways based on the observation
that the control group is often larger than the exposed group: imagine
the whole population versus coffee drinkers.
\begin{itemize}
\item[{\bf Balanced:}] We sample the control using a heuristic so to
  balance control and exposed, without changing the average {\em
    response} of control, doing so, we can compare directly the
  absolute number of visits but a relative measure is still
  preferred. Then generalize the result to the whole experiment.
\item[{\bf Unbalanced:}] We keep the size of each set unbalanced, as
  they are, and we compare their average response. There is no need to
  generalize the results found.
\end{itemize}
A balanced one will emphasize the actual response of each devices,
because their numbers are equal see Section \ref{sec:CMS} and Equation
\ref{eq:cms-lift}. This is natural we like to compare things directly,
one-to-one, but exposure touches a {\em small} set and we have to make
the control small. An unbalanced one may show how little contribution
the exposed group has in absolute number of visits, but because the
exposed and control groups are now different we must emphasize their
average performances. In principle, the balanced/unbalanced approach
should provide good estimates to the campaign performance especially
for large experiments. We shall show that the unbalanced approach is
general enough so that we can derive a version of the balanced
approach. In practice, there are constraints that will make the two
approaches different. Of course, there can be many variations and
applications, a complete comparison is beyond the scope of this work.
 

Assume that we have a tool to compute the {\bf response} of any user:
that is, for any device $d_i$ and any date $e_j$ we have a
quantitative measure
\[r(d_i,e_j)\]
that gives us the number of visits from time $e_j$ forwards and it
subtracts the number of visits before $e_i$. The response is a
difference between two interval of times in order to adjust for
features that are time sensitive (i.e., $e_j$); thus, we can account
for their effects (better). In social science and medical treatments,
a device owner can have a response for being exposed and a response
for being not exposed: a treatment $\gamma$ and not $\gamma$ (i..e.,
$r_{\not \gamma}$), sometimes placebo means no treatment and sometime
it means a different treatment, and both treatments could be given. In
such a scenario, the response would be this
\[r(d_i,e_j) = r_\gamma(d_i,e_j)-r_{\not \gamma}(d_i,e_j)\]
and we could estimate the effect of the campaign by estimating the
expectation of the response statistics:
\begin{equation}
L = E[r(d_i,e_j)] = E[r_{\gamma(e_i)}(d_i,e_j) - r_{{\not
      \gamma}(e_i)}(d_i,e_j)].
\end{equation} 
In practice, a user is either exposed or control. Thus for an exposed
device $r_{\not{\gamma}(e_i)}(d_i,e_j)$ is zero and for a control
device $r_\gamma(d_i,e_j)$ is zero, thus for control the response has
negative contribution. At the limit, the fact that exposed and control
can have different sizes is no issue with the expectation of L;
however, in practice the lift should be written as follows
\begin{equation}
\label{eq:lift-general}
Lift = E[r(d,t) | d \in \text{ Exposed}] + E[r(d,t) | d \in \text{
    Control}]. 
\end{equation} 
Where $E[x|y]$ is the conditional expectation of $x$ with respect to
$y$, the first mode of the response statistics, or lift, is a
comparison between expectations.  Considering that control response is
negative, lift is a difference in expectations.\footnote{An explicit
  difference $E[r_e]-E[r_c]$ is more common in literature as we report
  in the following.} First, we must consider correlation. Second,
expectations are computed by averages (i.e., if we use bootstraps,
several samples of averages). Equation \ref{eq:lift-general} is the
foundation of most comparative analysis in social science: our is yet
another social experiment and it can involve million of devices and
people across the country. In the following, we set the notations and
definitions, and we do our best in expressing Equation
\ref{eq:lift-general} on a clear mathematical footing. Our main goal
is the analysis of the statistics $r(d,t)$ and provide a clear and
complete characterization of it. We will dwell also in the lift
statistics which is the average of the original statistics.

We start with the definition of an {\bf impression}
$\iota(d_i,t,\ell)$: an impression has a device identification number
$d_i$, it has a time stamp $t$ in seconds, and it may have a
geographical locations as $\ell = (latitude, longitude)$. In practice,
an impression represents when a device is exposed to an advert and
possibly where this event happened.  We may distinguish two scenarios:
\begin{enumerate}
\item The impression belongs to a campaign we want to measure
  performance, thus the time is used to specify the first time a
  device is {\bf exposed} and
\item The impression has location $\ell$, we measure how close this
  device is to our location of interest.
\end{enumerate}
Intuitively, a campaign is a set of adverts delivered by different
means: mobile web, {\em apps}, sometimes using conventional digital
advertising such as websites' banners and sometime by TV commercials.
An exposed device has a {\em first time seen}. We can define it also
for a control device: it is when we have the fist recorded impression
during the campaign. Thus, all impressions after the first time seen
can be used for the measure of performance.

For us, the {\em goal of a campaign} is to invite the owner of a
device, a user, to visit a set of {\bf locations}. These locations are
the likes of coffee shops, department stores, or movie-theater show
rooms. Here, we define a location $l_i$ by its geographical location
$(lat, lon)$ and a campaign location set as $L_C$.

The first connection between impressions, locations, and campaign
performance, is by the definition of a {\bf hit}: a hit is an
impression with the following properties:

\begin{equation}
\label{eq:hit}
 h(d_i,t,\ell) =
  \begin{cases} 
   1 & \text{if } \exists l_i \in L_C, \exists K>0, \exists
   \|.\|:\ell\times\ell\rightarrow \R, \text{ s.t. } \|\ell,l_i\|<K
   \\ 
   0 & \text{otherwise }
  \end{cases}
\end{equation}

A hit is an impression arbitrarily close to one location. We intend to
use the $L_2$ norm thus the hint in the notation of $ \|.\|$, and it
is a distance function. Having a distance function is general enough
that, if we like, we can reduce the distance to a zero-one function as
to belong or not to any polygon describing the outside boundary of a
building. Eventually an altitude will be part of any geographical
location and more interesting distances will be used.

We shall digress a little by introducing a few important variations
enriching the definition of a hit. This is not required for the
understanding of our approach and it can be skipped at first
reading. After this digression we shall define a visit in Equation
\ref{eq:visit} on page \pageref{eq:visit}.
\newline

\mypar{Deterministic and stochastic hits}
In Equation \ref{eq:hit}, we say that an impression is a hit if it is
within a radius from our location of interest.  The radius is an
arbitrary choice and it represents a meaningful area.  This choice is
made a priori before the campaign started and by the client. This is
the first priority of being a hit and thus a visit.

Sometimes, the location is in a private parcel. This natural area is
described by a polygon or set of vertices. Once the parcel is
identified, any impression in the parcel can be a hit. This definition
is extremely useful for investigating locations with large
real-estate: that is, where the parking lot is as important as the
inside of the facilities. For example, this simple mechanism can
actually double the signal coming from hits and thus enhance the
experiment. We can envisions several combinations between the radius
and parcel based hit, these are beyond this short digression.

There is a third type of hit that is worth considering: it is a
stochastic hit. A hit with a range in $[0,1]$ can be liberating
because there are so many uncertainties that may contribute to a
different measure of distance between a location and an impression. If
we accept that the latitude and longitude of an impression inherit an
error, if we accept that our estimation of the earth radius is an
approximation that varies as function of the longitude, if we accept
that we may have just a sample of the impressions available, and
eventually that the users may have behavioral idiosyncrasies in their
use of the mobile phone. If we accept all of the above, then we can
move into a stochastic measure in order to relax the condition of {\em
  zero} hits.

This can be an error in measure, thus we could use a normal
distribution: we could argue that the distance is
\[ d +\epsilon \mbox{ where } \epsilon \in {\cal N}(0,\sigma_\ell)  
\]
Such an error could be used to create a gray area around the radius of
interest and thus smooth the hit response function accordingly. 

Another point of view is to consider the device path (i.e., the record
of impressions) as a stochastic process. A user visiting a location
must be entering the location and then leaving. If we measure an
impression just outside of the interest radius, what could be the
probability the device will step in unnoticed or it has stepped out
noticed?

With major arguments to be proven, a Brownian motion with a drift
describes our scenario nicely: we say that for the impressions just
outside the radius, their distance from the location is like  an inverse
Gaussian distribution.

\begin{equation} 
\label{eq:ig}
p_x(x|\mu,\lambda) = \sqrt{\frac{\lambda}{2\pi x^3}}\exp^{-\frac{\lambda(x-\mu)^2}{2x\mu^2}}
\end{equation}

For example, we take a location and for all impressions with distance
less than three time the radius, we compute the average and variance
\[ \dot\mu{=}\frac{1}{N}\sum_{i=1}^N d_i \mbox{ \bf and } S^2{=}\frac{1}{N-1}\sum_{i=1}^N (d_i -\dot\mu)^2 \]
Thus we could describe our process by
\[\dot d \in IG( \dot\mu,\frac{\dot\mu^3}{S^2})\] 
and a stochastic hit is based on a survival function
\begin{equation}
\label{eq:stocastic hit}
 sh(d_i,t,\ell) =
  \begin{cases} 
   {1 -CDF_{\dot  d}(d_i)} & \text{if } R < d_i  \leq R +\frac{R}{2}
   \\ 
   0 & \mbox{otherwise} 
  \end{cases}
\end{equation}
As for the parcel, stochastic hits are enrichment to the original
definition of hits, we let the data help describing the hits in
conjunction with external and accepted definition.  We shall present
experimental results and further discussions about this type of visits
in Section \ref{sec:brownian}.
\vspace{2cm}
 
We conclude here our digression and now we can define a visit from the
definition of hits.  In Equation \ref{eq:visit}, we may choose to lump
multiple hits in an interval of time into a single hit that we call
{\bf visit}:
\begin{equation}
  \label{eq:visit}
  v(d_i,t,\Delta t) =
  \begin{cases} 
    1 & \text{if } \exists t_i \in [t,t+\Delta t],i \in [0,N>0] ,
    \exists H>0, \exists \{\ell_m\} \\ 
    & \text{      \bf s.t. }  \sum_{j=0}^{N-1}
    h(d_i,t_j,\ell_m)>H \\ 0 & \text{otherwise }
  \end{cases}
\end{equation}
A visit is a function of a device, the locations in $L_C$ where we
have a hit, and the interval of time $\Delta t$. Such interval can be
arbitrary, for us it can be up to one day.

Now consider the flight of a campaign as a segment on a straight line:
there is a beginning and there is an end. A day is a single tick on
the segment and we have a discrete set of intervals or epochs $e_j$:
\begin{equation}
  \label{eq:visits}
  V(d_i,e_j) = \sum_j v(d_i,t_j,\Delta t_j) \text{ where }
  t_j,t_j+\Delta t_j \in [e_i, e_{i+1})
\end{equation}
Equation \ref{eq:visits} represents the {\bf visits per day} for the
device $d_i$ in the day specified by epoch $e_j$.  This is a time
series.  Given any epoch in this time line, we may want to compute a
discrete differential function to determine the grade of increase or
decrease. We call this {\bf response} and we use an {\em even}
symmetric weight function:
\begin{equation} 
  w_k = -w_{-(1+k)}, \forall k \geq 0   
\end{equation}
where 
\begin{equation}
  \label{eq:quasi-laplace}
  \begin{cases} 
    w_k={\bf e}^{-k}  & k \in [0,M-1] \\
    w_k=-{\bf e}^{k+1} & k \in [-M,-1]
  \end{cases}
\end{equation}
such as a modified Laplace function to weight accordingly visits at
different times.
\begin{equation}
  \label{eq:response}
  \Delta_M r(d_i,e_j) = \sum_{k=-M}^{M-1}V(d_i,e_{j+k})*w_k 
\end{equation}
The weight function is symmetric and the response $\Delta_M
r(d_i,e_j)$ follows. The present epoch $e_j$ has maximum weight
$w_0\sim 1$ as well as the previous day epoch $e_{j-1}$ ($w_{-1}\sim
1$ ). At a minimum, $e_{j-1}$ and $e_{j}$ summarize the discrete
difference of the visit rate; such a symmetric weight function will
help to smooth the response function. In practice, we introduce the
concept that a visit can belong to the interval $[0,1]$. Furthermore,
not every user has a response or visits, for that matter, at epoch
$e_j$. For example, if a device $d_i$ has its first time seen epoch
after $e_j$, it will not contribute. If the last impression is before
the epoch $e_{j-M}$, it will not contribute either.

Given a campaign $C$, its locations $l_C$, the flight beginning $e_1$
and its ending $e_T$ with $T>1$, and a set of devices $d_i$ with $i
\in [1,N]$, we call the set $D$; we may want to measure the
performance of the campaign by its expectation of the response; that
is, the average acceleration of visits.  A clear caveat is that not
all devices will be active and they will not contribute to the
response at every epoch: for each epoch $i$, $N_i$ of the $N$ will
contribute to the true acceleration; if we keep $M$ as a constant we
can omit the $\Delta_M$:
\begin{equation} 
  \label{eq:expectation}
  E[r(d,e)|(D,T)] \sim \frac{1}{T}\sum_{i=1}^T\Big[
    \frac{1}{N_i}\sum_{j=1}^{N_i}\Delta_M r(d_j,e_i)\Big]
\end{equation}
Thus, we can finally express our performance measure: We identify the
{\bf exposed group}, the set of devices that have been exposed to the
campaign adverts by {\bf E}, we find a suitable and comparable {\bf
  control group} that we identify by {\bf C}. Then the quality of the
campaign is summarized by the following expectation that we call {\bf
  lift}:
\begin{equation}
  \label{eq:lift}
  Lift = E[r(d,e)|(E,T)] -E[r(d,e)|(C,T)]
\end{equation}
The process summarized in Equation \ref{eq:lift} explains our original
definition of lift as in Equation \ref{eq:lift-general}, where in this
latter equations we show that time is common to both expectations and
the negative contribution of control is explicit because the response
is computed for exposed and control separately. However, our
formulation can be restated with little modifications if users have
multiple treatments, thus different responses, at different times.

\mypar{Practical considerations} The lift as expectation is powerful,
but it does not resonate to whom design the campaign. The sign is easy
to understand: there is a positive effect or there is a negative
effect. Often, we used to transform the lift measure into a relative
number, percentage:
\begin{equation}
  Lift_p = 100*\frac{E[r(d,e)|(E,T)] -E[r(d,e)|(C,T)]}{E[r(d,e)|(C,T)]}
\end{equation}
This is palatable because it provides a relative measure to a
reference baseline (i.e., control).  Unfortunately, for small control
lifts, the relative measure can be unbounded. A small campaign can be
unreasonably profitable/unprofitable and with large variances. In
general, the measure that appeals the most is a relative measure based
on the number of visits. For example, this is a simple normalization
that we use often
\begin{equation}
  Lift_v \sim 100*(lift*N*time)/(total visits)
\end{equation}
This is a relative measure of extra visits provided by the campaign.
We will hide under the hood which computation we use. In the
experimental section, we shall show relative lift (always based on the
expectation) and we use it to appeal the practitioner and for
presentation purpose: a projected measure in the range $[-100,+100]$
is more {\em pleasant} than the real measure in the range $[-10^{-6},
  10^{-6}]$.

\subsection{Exposure effect: First Time of Exposure and  Balanced Control} 
\label{sec:CMS}

By construction, we know the first time when we expose a device. This
is the analogous to the start of treatment in a clinical trial and it
is clear the importance of the response value $\Delta_M r(d_i,e_j)$
when $e_j$ is the first time seen (Equation \ref{eq:response}). In
practice, the $e_j$ is completely defined by the device and we can
describe this by $\xi(d_j)=e_i$. With a proper choice of the interval
$M$ and weight function, the exposed device contribution is limited to
the interval $[e_{j-M}, e_{j+M}]$. Thus Equation \ref{eq:expectation}
can be simplified to
\[ 
E[r | E] \sim \frac{1}{N}\sum_{j=1}^{N}\Delta_M r(d_j,\xi(d_j)=e_i)=\frac{1}{N}\sum_{j=1}^{N}\Delta_M r(d_j)
\]
Even thought it is not necessary, for symmetric purpose and for
computation purpose, we should find a similar epoch for the control
devices so that we can compute $E[r|C]$ in a similar fashion. We can infer an
epoch when the control device could have been exposed but we did not
expose them. Such a first time seen will be given in correlation to
the exposed devices and thus the computation can be carried on.  This
is easier if the exposed--control set is balanced, because we could
use random sampling to infer an average response for otherwise an
arbitrary choice of epoch.
\begin{align} 
  \label{eq:cms-lift}
  E[r|E]{-}E[r|C]  &  \sim\\
  & \frac{1}{N}\sum_{j=1}^{N}\Delta_M
  r(d_j,\xi(d_j)|E) -\frac{1}{N}\sum_{j=1}^{N}\Delta_M r(d_j,\xi(d_j)|C) \nonumber \\
  \label{eq:cms-lift-1}
  &  \frac{1}{N}\sum_{j=1}^{N}  \Delta_Mr(e_j) -  \Delta_Mr(c_j)\\
    \label{eq:cms-lift-2}
  & \sim E[r_e -r_c]
\end{align}

This formulation and the computation are appealing for a few and
important reasons.
\begin{enumerate} 
\item The response statistics $r_e -r_c$ is very intuitive and
  computationally appealing, it describes the original problem
  clearly.
\item If the control set is independent of the exposed set and we have
  a good estimate of the first time seen for both, this will be a well
  formulated experiment. Thus we can do matching to remove targeting
  bias and compare performance by slicing the data accordingly. We can
  compare the original lift and the matched lift.

\item The computation of lift as average is a sound estimate of the
  expectation and thus of the campaign lift.

\item Interesting constraints can be applied to expose and control
  together: for example, we could count only visits that happen in a
  short interval after exposure and thus reject long term effects.

\item It is like 1-1 matching is applied already.

\item The lift measure is an expectation of the difference of two
  functions and not the difference of expectations; thus a single
  variance can be associated instead of two and their correlation
  matrix.
\end{enumerate} 
From a different prospective, these advantages are hindrances. For
example, loosing the relation between visits and epoch may be too much
of a simplification. We may want to compute explicitly:
\begin{equation}
  \label{eq:lift-time}
   \frac{1}{N_i}\sum_{j=1}^{N_i}\Delta_M r(d_j,e_i)
\end{equation}
In Equation \ref{eq:lift-time}, we represent a time series of the
visits acceleration and we show how a campaign is effective to the
granularity of the epoch of $e_i$. Also, exposure should have long
term effects because exposure is a continuous process limited only by
a frequency cap.

The estimate of first time seen for control devices is a fragile
process especially for long campaigns and large control set: force a
control user to be measured at an arbitrary date in the process may
not represent its average behavior. Lift as expectation and variance
as single number is very powerful: it comes with the condition that
the implicit pairing in Equation \ref{eq:cms-lift-2} is proper, but
matching will come only after this pairing, which could be too late.

Nonetheless, we shall show experimental results using this balanced
approach.

\subsection{Unbalanced Control: A visit based choice.}
\label{sec:control-selection}
Let us repeat here what is the goal of a campaign: it is an invitation
to visit a set of locations and the exposed group is given. However,
just {\em seeing} one location is another invitation to visit and this
is an invitation directed to any one nearby. Clearly, a close distance
is of the utmost importance for both exposed and control and this is
obviously true for national campaigns towards local enterprises: for
example, {\em Blue Bottle Coffee} has locations in Tokyo Japan, San
Francisco, Palo Alto, Oakland, and New York and a connoisseurs may
visits all them during the span of a week, but someone living in Los
Angeles is neither a great target for exposure nor a great example for
control.

First, we can draw a circle of one mile around a coffee shop and
suggest that the people in this circle will visit our location.
Considering that who visits will be eventually in the circle, this
seals a strong group and, in it, there will be part of our exposed
group. The connoisseurs will be there (if we have impressions for
them, yes even in Tokyo) but also any passer by. 

Given this exposed--control set, we can compute the lift for the
exposed and for the control. These sets will have all the visits but
they will not represent the whole exposed population and will not have
enough similarities to compare to each other (i.e., matching). We may
have to increase the circle to two miles to capture twice as many
exposed and twice as many control. Notice, control will be much larger
than exposed unless we saturate the area and the population
(economically not smart). We need to have enough users so that we can
do a meaningful matching and thus removing any bias in the sample:
this means, when most exposed can be matched with control and
viceversa.

\section{Audience Specification: User Profile and Location Graph}
\label{sec:keywords}

In the following section, we shall describe an original and
deterministic approach to create a feature space describing any user
and any location by a set of keywords, which are key--value pairs and
the values are probabilities in the continuous space in $[0,1]$. For
example, we may introduce a probability for a user to be {\em
  female}. We shall explain our inference process but intuitively a
user visiting a location will inherit some of the features of the
location and, in turn, the location will do the same. First, a
business location attracts an audience; Second, its neighbors will
inherit some of this audience; Third, a lot or a few people may visit
a business location, and Fourth, who is visiting also composes the
audience of the business. All these interactions will affect the
features of both businesses and visitors.  We shall start describing
the location graph in Section \ref{sec:location-graph} and the user
profile in Section \ref{sec:user-profile}.

\subsection{Location Graph}
\label{sec:location-graph}

In general, a triplet $(pid_i,lat_i,lon_i)$ defines a location $l_i$
and it may represent either a business location or a census-based
location.  Given two locations, we can compute different type of
distances: Haversine formula, Manhattan, or a version of
$L_1$. Independently, we denote the distance between two locations as
$d_R(l_i,l_j).$ We use an estimate of the Earth radius in line with
the literature and, of course, the smaller is the real distance the
better is the approximation.

Consider a ${\bf cell}_{i,j}$ defined by point $(lat_{i},lon_{j})$ and
point $(lat_{i}+\Delta x,lon_{j}+\Delta y)$, where $\Delta x$ and
$\Delta y$ are arbitrary positive constants.  The point
$(lat_{i},lon_{j})$ describes a rectangular cell. We define as cell
any $(lat,lon) \in {\bf cell}_{i,j}$ so that 
\[ lat_{i}\leq lat<
lat_{i}+\Delta x \text{ \bf and } lon_{j} \leq lon< lon_{j}+\Delta
y.\] Assume we can draw a cell {\bf partition} for any geographical
area (i.e., the United States): that is, any location is in one cell
and any two cells have empty intersection.  Any ${\bf cell}_{i,j}$ has
eight neighbors specified by the following points in counter clockwise
fashion {${\cal N}({\bf cell}_{i,j})$}:
\[
\begin{array}{ll}
{\bf cell}_{i-1,j-1}&=(lat_{i}-\Delta x,lon_{i}-\Delta y), \\
{\bf cell}_{i-1,j}  &=(lat_{i}-\Delta x,lon_{i}),          \\ 
{\bf cell}_{i-1,j+1}&=(lat_{i}-\Delta x,lon_{i}+\Delta y), \\ 
{\bf cell}_{i,j+1}  &=(lat_{i},lon_{i}+\Delta y),          \\ 
{\bf cell}_{i+1,j+1}&=(lat_{i}+\Delta x,lon_{i}+\Delta y), \\ 
{\bf cell}_{i+1,j}  &=(lat_{i}+\Delta x,lon_{i}),           \\ 
{\bf cell}_{i+1,j-1}&=(lat_{i}+\Delta x,lon_{i}-\Delta y),  \\
{\bf cell}_{i,j-1}  &=(lat_{i},lon_{i}+\Delta y).           \\
\end{array}
\]
With the concept of distance, we know that any two locations $l_m$ and
$l_n$ in ${\bf cell}_{i,j}$ must have $0\leq d_R(l_m,l_n) \leq
d_R((lat_{i},lon_{j}),(lat_{i}+\Delta x,lon_{j}+\Delta y))=D$. Thus if
one location in ${\bf cell}_{i,j}$ has a neighbor at distance no
farther than $D$ then it has to be in any of the 8 neighbor cells
above. Now that we have defined distance and cell partitions, we can
define and compute a geographically distributed {\bf location graph}.

\singlefigure{0.6}{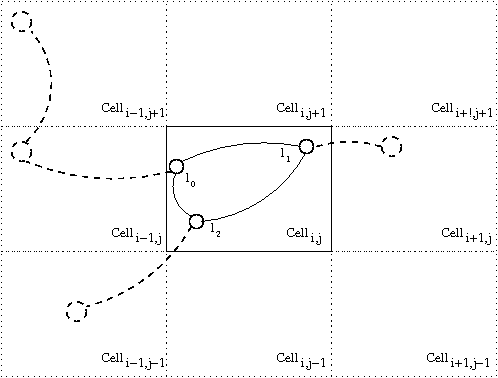}{Example of geographical distributed location
  graph}{fig:cell} 

Assume we set the constant $\Delta x = \Delta y = \Delta$ and we have
a cell partition. For any cell ${\bf cell}_{i,j}$, we collect all the
locations in ${\bf cell}_{i,j} \cup {\cal N}({\bf cell}_{i,j})$. For
every location $l_m \in {\bf cell}_{i,j}$, we compute a distance with
all other locations $l_n\neq l_m \in {\bf cell}_{i,j} \cup {\cal
  N}({\bf cell}_{i,j})$. We then create a node in a graph associated
with $l_m$ where we store all the neighbors information and their
distances (edges) if the distance is less than $D/2$ (say).  The graph
has a geographical key ${\bf cell}_{i,j}$ and each location has
information about its first degree neighbors, which must be in ${\bf
  cell}_{i,j} \cup {\cal N}({\bf cell}_{i,j})$. In Figure
\ref{fig:cell}, we show an example of location graph. The cell
partition describes a grid and we shall introduce grid algorithms. In
the following section, we shall explain briefly the graph building
computation, which is the basis for all our graph computations.

\subsection{Location Graph: Distance Computation}
\label{sec:distance} 
For simplicity, we have a injective function that map any location
$l_m$ to only one cell, 
\[ J(l_m)={\bf cell }_{i,j}, \]  
and given a ${\bf cell }_{i,j}$ we can compute ${\cal N}({\bf
  cell}_{i,j})$.  The computation of the location graph follows and it
is the foundation for any graph algorithms in this work:

\begin{itemize}
\item[{\bf Broadcast:}] For every location $l_m$ in the graph, we
  compute the cells ${\cal N}(J(l_m))$, and we broadcast $l_m$ to the
  them. In practice, the location $l_m$ is associated with a node in
  the graph.

\item[{\bf Computation:}] For every cell ${\bf cell}_{i,j}$ and for every pair
  $l_s\neq l_t$ in the cell, we compute all $d_R(l_s,l_t)$ distances
  and we create an edge between the nodes with their distance. 

\item[{\bf Reduce:}] For every cell ${\bf cell}_{i,j}$, we store the
  nodes $l_s$ so that ${\bf cell}_{i,j} = J(l_s)$
\end{itemize}

The location graph is a hierarchical graph, a cell describes a
circumscribed area where the graph has locations with connections
within the cell and, possibly, only to its neighbors cells. A location
$l_i=(pid_i,lat_i,lon_i)$ identifies all its direct neighbors
$\pi(l_i)$, its own cell $J(l_i)$, and the neighbors' cells ${\cal
  N}(J(l_i))$. If a location has also an altitude, think the Dubai's
tower, the distance can be enhanced in order to distinguish locations
beyond latitude and longitude but there is no changes about cells and
neighbors.

\subsection{Location Graph: Keywords} 
\label{sec:prior} 

We are ready to dwell into keywords and their notations. Consider a
location $l_{j}$ where we enumerate the locations using an integer
$\ell \in [1,J]$. In the same way, we enumerate the keywords using an
integer in $k \in [1,K]$. A keyword is associated to a step $s$; this
has two meanings. First, $s$ represents the time in a time series;
Second, at any step $s$ we perform a computation among only direct
neighbors: thus from step $s-1$ to step $s$, each step propagates
keywords values in the graph by direct connections and thus we
propagate keywords to {\em two level} connected neighbors. The last
and important identification of a keywords is the category where this
feature is derived from, we enumerate the categories by $c \in [1,3]$:
First, from its direct neighbors; Second, weighted by number of visits
and Third, the keywords from visitors.  We shall clarify these
distinctions as soon as we express the keywords and their
applications. In short, we can represent any keyword by
\[\pred^c_kw^s_\ell.\] 
As reinforcement, we have: $\ell$ location, $s$ step, $k$ keyword and
$c$ category.

Using matrices and using notations to describe each location $\ell$,
our computation is an exponential smoothing, we have $\Vc{W}^s_{\ell}
\in \R^{3\times K} $:
\begin{equation}
\Vc{W}^s_{\ell} = \left[ \begin{array}{c}
\Vc{n}^s_\ell \\ 
\Vc{v}^s_\ell \\
\Vc{u}^s_\ell  
\end{array} \right]
 \Leftarrow {\bf \Lambda}\oplus(\Vc{W}^{s-1}_\ell)+({\bf
  I-\Lambda})\oplus \f(\X)
\label{eq:exponential-smoothing}
\end{equation}
We shall specify each term by unfolding the matrix notation completely
(the uncommon symbols $\oplus$) and we shall shed lights on the use of
exponential smoothing.

\subsubsection{Neighbors $c=1$} In real estate, the location and its 
neighbors are key to any property value.  Given a location $\ell$ and
its neighbors up to $|\pi(\ell)|=N_\ell$ at step $s$, we compute the
update keyword value by the following exponential smoothing:
\begin{equation}
  \Vc{n}^s_\ell = \Vc{\lambda}*\Vc{n}^{s-1}_\ell +
  \frac{(\Vc{1}-\Vc{\lambda})D_\ell }{N_\ell}* \sum_{j \in
    \pi(\ell)}\frac{\Vc{n}^{s-1}_j}{1+d(j,\ell)}
\label{eq:neighbors}
\end{equation}
Where we have 
\[\Vc{a}* \Vc{b} =[a_1b_1, a_2b_2, ..,  a_Kb_K]^t, \] 
we identify the transpose of a vector $\Vc{v}$ with the usual
superscript $\Vc{v}^t$, the scalar by vector follows the usual rules
\[a\Vc{v} = [av_1, av_2,  .., av_K]^t,\]  
\[ (\Vc{1}-\Vc{\lambda}) = [1-\lambda_1,..,1-\lambda_K]^t \]  and 
\[ D_\ell= \sum_{j \in \pi(\ell)} \frac{1}{1+d(j,\ell)}\] and $d()$ is
our distance function. 

The notation tries to keep an intuitive format but we understand the
difficulties associated with it.  Here is our interpretation of
Equation \ref{eq:neighbors}. We compute the average of the keywords of
the neighbors; each neighbor contributes as a linear and decreasing
function of its distance from the location: that is, the closest the
distance is, the largest contribution will be; intuitively it is the
inverse of a cost and the cost is the time required to go from one
location to the other; we assume they will provide information to the
location $\ell$; if $\ell$ has no other source and the neighbor
keywords do not change, then at steady state $\Vc{n}_\ell$ will
converge to the average of its neighbors.

\subsubsection{By number of visits, $c=2$} Given a location $\ell$, we can
compute the number of visits to this location in between $s{-}1$ and
$s$, and we identify this count by $V_\ell^s$. Thus we express the update
as follows

\begin{equation}
\Vc{v}^s_\ell= 
\frac{V^{s}_\ell}{\Upsilon_\ell}(\Vc{\mu}*\Vc{v}^{s-1}_\ell) 
+ \frac{(\Vc{1}-\Vc{\mu})}{N_\ell}*
\sum_{j \in \pi(\ell)} \frac{V^{s}_j}{\Upsilon_\ell} \Vc{v}^{s-1}_j 
\label{eq:visits2}
\end{equation}
Where $\Upsilon_\ell$ is $\sum_{j \in \pi(\ell)} V_j^s$. In practice,
a location with a lot of visits will dominate the surrounding
audience. By construction, the neighbors locations are relatively
close and thus inherit these hot spots audience. In practice, a few
business place themselves in close proximity to others to gain access
to their audiences.

\subsubsection{By visit types, $c=3$} Assume we account for the keywords of
the visitors of location $\ell$; for example, we create a distribution
of the visitor keywords from epoch $s{-1}$ to $s$ and to show that is
related to the computation of the keywords we use the following
notation $\bar{\Vc{u}}_\ell$, then we count the number of visitors
$V_\ell^s$ as above.

\begin{equation}
\Vc{u}^s_\ell = \Vc{\nu}*\Vc{u}^{s-1}_\ell + 
\frac{(\Vc{1}-\Vc{\nu}) }{\Upsilon_\ell (N_\ell+1)} * \sum_{j
  \in \pi(\ell) \cup \ell}V_j^s\bar{\Vc{u}}^{t}_\ell 
\label{eq:visits3}
\end{equation}
The computation by visit type combines the traffic of all direct
neighbors and create a weighted average. For example, we can emphasize
the direct visitors instead of neighbors'. This is one approach to
harness keywords associated to users, it is independent from the
graph, and create a ripple effect of otherwise sparse and rare events
such as visits.

\subsubsection{All together} We foresee the case of other dimensions and
categories that can be added to the previous ones, but here we explain
how we summarize the keywords weight. In summary, $\Vc{n}^s$
represents the graph contribution and it changes only when the graph
changes; $\Vc{v}^s$ introduces the idea that neighbors have different
contribution as a function to the number of visitors; $\Vc{u}^s$
represent the contribution from each visitors to each neighbors
locations.

We combine them as independent

\begin{equation}
 \Vc{w}^s_\ell = \Vc{\Gamma}^t  \left[ \begin{array}{ccc}
\Vc{n}^s_l &
\Vc{v}^s_l &
\Vc{u}^s_l  
\end{array} \right] = \Vc{\gamma}^n*\Vc{n}^s_l +\Vc{\gamma}^v*\Vc{v}^s_l + \Vc{\gamma}^u*\Vc{u}^s_l
\label{eq:gammas}
\end{equation}

where 
\[
\gamma^n_k = \begin{cases} 0 & [\Vc{n}^s_l]_k = 0 \\ 1/\text{number of
    non zeros } [\Vc{{u/v/s}}^s_l]_k& \text{otherwise}
\end{cases}
\]
In practice, we compute an simple average for each keywords but if any
dimension provide no keyword entry, that is zero, we do not account
for them. In this way, when we will add more dimensions, they will
enrich the keywords, {\em density}, and they will not increase their
value.

\subsubsection{Priors, distributions, and probabilities} 
Sometimes, a business knows the audience they have or they must
have. This is a priori information or simply prior. At first, it may
seem difficult to envision businesses with a specific and limited
audience; however, in practice, there are gender-specific services
such as Obstetrics and Gynecology, there are age-specific business
such as liquor stores, there are income-specific ZIP codes (upper east
side) or business such as Ferrari dealers.

In this prior category, we add averages as well: for example, by
Census or other means, we know the {\em averages} about the population
of a ZIP code (thus a location or locations). Such averages help to
narrow down the most likely features. In fact, we use probability as
estimate of the feature distribution and use these as prior
information. For example, we may have information about a ZIP+PLUS4
area, which we represent by its centroid latitude and longitude, the
locations in this ZIP will be either connected directly to it or a few
step way, and thus the ZIP will enrich their own keywords by
proximity. Of course, an area into a single node in the location graph
is an approximation but it is one simple way to feed the graph with
priors otherwise not available.

Prior features are set and no further computation is needed. There is
only one exception introduced to cope with imprecise and erroneous
data: we normalize the keyword values in order to keep meaningful
distributions, this may require scaling the keyword values.

\subsection{Location Graph: Update}
\label{sec:update}
Once we have a location graph with its features, we may have to rebuild
the graph because a few businesses relocated, new businesses started,
and old business exited.

In the graph, a business exiting translates into a node deletion and
thus edges deletions to all its neighbors. Any modification to the
list of locations will require a graph update. In this scenario, the
new graph will inherit the keywords from the old graph but we must
propagate the effect of the new connection by updating the keywords
accordingly to Equation \ref{eq:exponential-smoothing}.

\subsection{Location Graph: Iterative Algorithm and cycles}
\label{sec:iterative-algorithms}
We can iterate the computation of keywords in order to collect signal
from locations father away in the graph or just to adjust the keywords
value because of graph modifications.

\begin{itemize}
\item[{\bf Broadcast:}] For every location $l_m$ in the graph, we
  determine ${\cal N}(J(l_m))$, and we broadcast $l_m$.

\item[{\bf Computation:}] For every cell ${\bf cell}_{i,j}$ and for
  every $l_s$ in the cell, we compute Equation
  \ref{eq:exponential-smoothing}--\ref{eq:gammas}

\item[{\bf Reduce:}] For every ${\bf cell}_{i,j}$, we store the
  nodes $l_s$ so that ${\bf cell}_{i,j} = J(l_s)$
\end{itemize}

This is an iterative algorithm that will propagate the keyword values
across the graph. In this scenario, the exponential smoothing
introduces yet another dimension: it will reduce, edge by edge, the
effect of cycles in the graph. In practice, one degree neighbor will
contribute $(1-\lambda)$, two degree neighbors will contribute
$(1-\lambda)^2$ (i.e., $i$-degree will provide $(1-\lambda)^i$). In
practice $(1-\lambda)\sim 0.5$ and thus the effect will be
$\frac{1}{2^i}$, after three levels the contribution is negligible
with respect to the local nodes and its direct neighbors. Nonetheless
the smoothing can be tuned for the keyword and for the dimension.

\subsection{User Profile}
\label{sec:user-profile}
In the previous section, we show that if we have information about the
users visiting a location, then we can enrich the location and its
neighbors. In this section we show that we can also take the location
information and enrich its visitors.

Assume we have already a set of users $\Vc{V}^{s-1}$ at step $s-1$
with their keywords and we want to compute the next step. We enumerate
the users $n\in [1,N]$ and we specify the users by the same keywords
as for the location graph: $\Vc{v}_n$ is a vector of keywords.  We
have at our disposal the following information: Prior ($\Vc{P}^s$),
Visits, and graph locations. The user profile computation has the
following formulation:
\begin{equation}
\Vc{V}^{s}= \Vc{\rho}* \F(\Vc{V}^{s-1},\Vc{P}^s)) + (\Vc{1}
-\Vc{\rho})*\Vc{X}^s.
\label{eq:users}
\end{equation}
We shall expand and clarify all terms and how we compute
visits (briefly explained and used in Equation \ref{eq:visits}). The
order of the computation is important and the evaluation goes
naturally from left to right because priors have priority.

\subsubsection{Visits and location graph $\Vc{X}^s$} 
From step $s-1$ to $s$ we gather the foot print of users. Listening to
real time bidding, we can observe a sample of impressions:
geographical location, time, and keywords associated to the users for
an advert; also we can collect similar information from third parties
pixels. These two sources do not overlap. The interval from $s-1$ to
$s$ represents an interval of time of one week. This is a geographical
distribution of impressions: $G^s$ for short.

We can gather all user information and thus average the keywords in
different geographical locations and create a distributions of the
keywords: we describes this information by $\Vc{{f}}^{s}_n \leftarrow
G^s$.

We merge the location graph at step $s-1$ with the $G^s$. We can
compute the user visits with respect the location graph and we can
collect the keywords from the location visited: $\Vc{g}^{s}_n$. Thus,
we have
\begin{equation}
\Vc{x}^{s}_n= \Vc{\gamma}^t_f * \Vc{{f}}^{s}_n +
\Vc{\gamma}^t_g* \Vc{{g}}^{s}_n
\end{equation}

\subsubsection{Priors} In every interval from $s-1$ to $s$, we collect
registration information, that are voluntary information the users
provide to a carrier, phone application, and others. This information
will change and it will describe the current users of the
device. Priors are so important that we can use them stand alone and
create discrete classes. This finds application to the balanced
approach as we shall describe in Section \ref{sec:experiments}.

In the contest of applying priors to user keywords, we must take care
of any inconsistencies with the previous priors:
\[ 
\bar{\Vc{V}}^{s-1} = \F(\Vc{V}^{s-1},\Vc{P}^s))
\]
this computation assesses the effect of the current Prior and we
update the user profile accordingly: a simple example is when the user
suggested to be a {\em female} and now to be a {\em male}. If left
unattended, this user will belong to male and female category.  In
this scenario, current priors will not substitute the past prior, it
will clear both genders and we will be open to suggestions from their
visits patterns, opening the opportunity of a probability of gender
based on their foot traffic.

As for the location graph, priors are unchangeable, we consider them
as simple truths. Thus, Equation \ref{eq:users} involves no
computation using priors. However, as for the location graph, we
perform an adjustment in order to keep keywords distributions
consistent.

This concludes the description of how we compute the user features. As
a summary, we compute features for locations and users as a function
of time and their interactions. Now, we shall describe how we use
these feature for matching, that is the art of compensating the
original bias introduced into the exposed group.

\section{Matching}
\label{sec:matching-intro}
To describe the efficiency of a campaign, we compute the variance and
the expectation of the response statistics, that is, the lift as in
Equation \ref{eq:lift-general}:
\begin{equation}
\label{eq:lift-general-2}
Lift = E[r(d,t) | d \in E] + E[r(d,t) | d \in C]
\end{equation} 
With the term quasi experiments, we refer our inability to choose the
exposed and the control group before the experiment as in a randomized
clinical experiment. However, we can still choose how the control
group interacts with the exposed group by an appropriate selection: in
practice, we sample the universe and we find for every device $d_i$ a
corresponding ${d'}_i$ in the control and the best we can do is to
compute the following,:
\begin{equation}
E_d\big(E(r_\gamma|d,E) - E(r_{\not\gamma}|d',C)\big)
\label{eq:2}
\end{equation}
where we represent conditional expectation as usual, $r_{\gamma}$ is
the response to exposure, and $r_{\not\gamma}$ the response to not
exposure, from the original work and notation.  We notice that the
notation infers a matching with suitable pairing for every device $d$ a
device $d'$. In the following, we shall provide an interpretation of
Equation \ref{eq:2}. Notice that Equation \ref{eq:lift-general-2}
and \ref{eq:2} are equivalent if the choice of exposure (i.e.,
$z(d_i)=1$ if $d_i$ in Exposed) is {\em strongly
ignorable}, \cite{RosenbaumR1983}. This is clear by using the original
notations:
\begin{equation}
(r|E,r|C) \orthogonal z|d  
\end{equation}
For every person in the sample $d_i$, the fact that $d_i$ is exposed
or not, is orthogonal to the value of the response.  As Dawid's
stated \cite{Dawid1979} $X \orthogonal Y$ where $X$ and $Y$ are random
variables, then
\begin{enumerate}
   \item $P(X=x,Y=y) = P(X=x)P(Y=y)$, 
   \item $P(x|y)  =P(x)$, 
   \item $P(x,y) = a(x)b(y)$, and 
   \item $P(x|y) = a(x)$.
\end{enumerate} 
The second case is the most practical for us: the distribution of the
response is independent of the distribution of the exposure
selection. In practice, we call general lift the result of
Equation \ref{eq:lift-general-2}. We do not expect targeting to be
strongly ignorable.

This is not just a mathematical tool; Equation \ref{eq:2} suggests
that exposed and control must be paired equally. This has a natural
application to the methodology described in Section \ref{sec:CMS}
where the exposed-control pair is balanced and matching will provide
an estimate of Equation \ref{eq:2} by 1-1 matching. In general, a good
control can be large, the computation of Equation \ref{eq:2}
difficult, and targeting specific and not ignorable, then we must
estimate something like the following instead:
\begin{equation}
\label{eq:lift-general-3}
E[r_\gamma(d,t) | d \in \text{ E and } M(d) \in \Omega] - 
E[r_{\not \gamma}(d,t) | d \in \text{ C and } M(d) \in \Omega ]
\end{equation} 
where $M()$ is an appropriate function and $\Omega =
[M(d):\text{E}\rightarrow \R] \cap [M(d): \text{C}\rightarrow \R]$, is the
intersection of the projection of exposed and control into the image
of $M$. This means we are changing the response statistics by
filtering, clustering, and weighting within the cluster; then we
compute expectations for the average lift. The
Equation \ref{eq:lift-general-3} provides an insight to the
Equation \ref{eq:lift-general-2} and, in general, there is no
equality.

\subsection{Matching: Previous Work}
\label{sec:previous-work}

The seminal work by Dawid \cite{Dawid1979} introduces the notation and
the formalism of {\em ignorable}, the main property to make a quasi
experiment closer to a random experiment.  

The population composing our experiments has explanation features, see
Section \ref{sec:keywords}.  The feature space is a multidimensional
space: multivariate. As the features can be used to discriminate the
exposed group from the control, we can use them to build $M()$ of
Equation \ref{eq:lift-general-3}.  The author
in \cite{Rubin1976c,Rubin1976b} introduced examples thus paving the
path for matching using propensity score. The authors in
\cite{RosenbaumR1983} developed the theory connecting the importance
of the propensity scores to matching. The propensity score transforms
a multidimensional space problem, which users have the closest set of
features, into a one-dimensional problem, which users have the closest
score number.

Assume that $\Vc{x}_i$ is a feature vector for the person $i$; we
estimate $z_i$ as $\zeta(\Vc{x}_i)$ in the following
Equation \ref{eq:linearmodel}, in turn we may use a linear model such
as in \cite{McCullaghN1989}, Chapter 4.3:
\begin{equation} 
\log\frac{\zeta(\Vc{x})}{1-\zeta(\Vc{x})} \sim \beta_0 + \sum_{i=1}^K\beta_ix_i
\label{eq:linearmodel}
\end{equation}

Rosenbaum and Rubin explain how any function based on $\zeta(\Vc{x})$
or, for that matter, its linear model in
Equation \ref{eq:linearmodel}, can be used for the computation of the
matching. We can use it in Equation
\ref{eq:2}  and we can determine $M()$ in Equation \ref{eq:lift-general-3}.  
Compute the model and determine the $\dot{\beta}$.  Take two people
from the experiment with their feature set $\Vc{x}_i$ and
$\Vc{x}_j$. If $\Vc{x}_i=\Vc{x}_j$,
$\zeta(\Vc{x}_i)=\zeta(\Vc{x}_j)$. Also, thanks to the linearity and
continuity of the estimation, $|\Vc{x}_i - \Vc{x}_j|<\epsilon $
translates to $|\zeta(\Vc{x}_i) -\zeta(\Vc{x}_j)|< \delta$ with
$\epsilon,\delta>0$ and arbitrary small, whether or not $z_i=z_j$.

The work by the authors in \cite{DehejiaW1998} is probably the most
cited attempt to provide a first evaluation of all the matching
algorithms using propensity scores. As of today, this work is a great
beginning and most of the current libraries implementing matching
provide the data from this reference (see
\cite{HansenK2006,IacusKP2012,Griffin2014,HoImaKin07,Sekhon2011,SekhonG2012}).  For a better view of the research, please consult the
references in \cite{MorganW2007,GuoF2015}.

Often the estimate of the propensity scores provides only a handful of
values, discrete, and thus classes with very large gaps. By
construction and in this scenario, any matching will be balanced and
thus it does not provide any meaningful measure of quality. Also it
means that the matching is many-to-many. Also, assuming we have to
match one exposed user with one control user, eventually we have to
sample either one. We may believe that random sampling should not
change the average behavior; however, in practice, when the events
are rare, sampling will remove them further and actually affect the
final results. This simple consideration explains why it is difficult
computing Equation \ref{eq:2} and also undermines balanced matching
based on $\zeta(\Vc{x})$, which is unrelated to the response.

In this work, we present a single algorithm that will work with both
propensity score and clustering algorithm. Also we show the danger of
perfect matching while estimating the exposed effect in
Equation \ref{eq:lift-general-3}. To overcome this problem, we
introduce a novel quality measure of the matching algorithm and an
un-balanced algorithm where the response distribution is not known a
priori and thus it is not affected by matching.

\subsection{The algorithms}
\label{sec:algorithm}
The propensity score describes a multidimensional vector by a
probability or a score with a continuity property stemming from
Equation \ref{eq:linearmodel}. The propensity model for $N$ users with
$K$ features is computed by an iterative algorithm with $J$ iterations
computing a $QR$-factorization in each step: $O(JKN^2)$. For us $K=25$
and $J<10$. Thus $O(N^2)$ is a good approximation of the
complexity but it is often executed in native and fast code.

The other property is that the scores can be sorted and a caliper can
be introduced naturally using the strong order of the scores. Sorting
$N$ scores takes $O(N\log N)$ and thus creating the matches with any
caliper takes an extra $O(N)$, a further pass. Optimal algorithms will
circumvent calipers and try to find the closest element independently.
This takes $O(\frac{N^2}{2})$ comparisons; for example, this is the
default implementation in \cite{Sekhon2011}. For large $N$ a full
comparison is not efficient \footnote{Even though the complexity
$O(N^2)$ is the same as the QR-factorization, this is not highly
optimized, the constant factor if much larger than 250 and it is often
not acceptable for $N>10^6$. } and sorting is by far the better
solution.

Complexity is one problem.  The other problem is sampling due to
matching: when exposed and control have different number of users and
we match one-exposed-user-to-one-control-user, we sample the larger
one. If the events we measure in the response have a distribution with
long (fat) tails such as in stable
distributions \cite{Mandelbrot:1960:PLL}, sampling will censor rare
but important events. In general, we do not want to use the response
for matching, to avoid bias; however, the change in the response
distribution can be used as a measure of quality: for example, we can
provide distance measure and its confidence using several
methods \cite{DAlbertoD2009}. Here, we use a simple, consistent, and
intuitive measure: entropy \cite{Shannon1948}. Small entropy changes
for exposed and control before and after matching mean a
representative matching. To achieve this entropy balance, we may opt
for unbalanced matching and get away from one-to-one matching. We
shall explain how we use weights and how they affect the computation
of Equation \ref{eq:lift-general-3}

\subsection{The Idea}
\label{sec:idea}

The propensity score defines clusters; that is, users with the same
score are in the same cluster. If we sort these scores, then we
sort/organize the clusters. By construction, consecutive scores in the
sorted list represent close clusters thanks to the continuity property
of the propensity score.

In practice, we score each user in the corpora composed of both
exposed and control. We sort the corpora by the score; then we apply
the following algorithm:
{ \small
\begin{verbatim} 
 1: i = 1
 2: scoreR = Corpora.scores[i]
 3: i = i +1  
 4: while (i<length(Corpora)) {
 5:  score = Corpora.scores[i]
 6:  if (scoreR != score) { # New Cluster
 7:    TE = length(tmpE) # Exposed 
 8:    TC = length(tmpC) # Control
10:    if (TE > 1 && TC>1)  { # Non Empty cluster 
11:      N = max( TE, TC)
12:      if (BALANCED) # or further random sample 
13:        U = union(tmpE[1:N],tmpC[1:N]) 
14:      else 
15:        U = union(tmpE,tmpC)
16:      matches = union(matches, U)
17:    } 
18:    tmpE =c(); te = 0
19:    tmpC = c(); tc = 0
20:    scoreR = score;
21:  } else  {
22:    if (Corpora.exposed[i]==1)    tmpE[te++] = i 
23:    else                          tmpC[tc++] = i 
24:  } 
25:  i = i+1
26: }
27: Corpora = Corpora[matches] # Matching done
\end{verbatim}
}

The pseudo code above matches the exposed and control by clusters.  We
can choose to have a balanced matching or unbalanced. Notice that
there is no matching in between clusters: exposed or control coming
from different clusters will have different scores. It is actually
easy to encompass this problem by moving the assignments at lines 18
and 19 inside the loop and at the end of the condition at line 10 (we
call this Caliper activation and it makes sense only for
propensity-score algorithms).

In practice, we could use any other means to cluster the users. For
example, if we use any $k$-mean clustering algorithms, the scores are
the cluster labels and the algorithm above will not change.  The
matching is actually decoupled by the scores: propensity score and
$k$-mean algorithms can be applied in combination. For example, we use
propensity score first with adaptive calipers to estimate the number
of clusters and, then optionally, we apply $k$-mean algorithm.

In practice, we compute the propensity score by generalized linear
model (GLM) \cite{McCullaghN1989}  and the $k$-mean by
\cite{Forgy1965,Hartigan1979}. Both methodologies are well known and
available. From our side, we would like to expose the matching
algorithm as a computational kernel and thus apply it to large
problems.

\subsection{Quality Measure and Confidence}
\label{sec:quality}

We are aware of several matching algorithms: {\em exact, cem,
  subclass, nearest, genetic, full, and optimal}. The list is
  longer. Our implementation falls among the first four.  This
  scenario begs for a simple question: how can we compare the results
  of matching. The package {\em MatchIt}, which offers an interface to
  all of the above algorithms, takes a step back and lets the user
  decide the statistics about the matched sets, even the computation
  of the lift is avoided.

Often the quality of matching is based on the estimate of the variance
of some sort.

\mypar{Sampling a normal distributed corpora} Each user has a response
$r_i$ (i.e., $\Delta_M r(d_j,e_i)$ in Equation \ref{eq:expectation}),
with $N$ users, we can measure the average
\[ \mu_N =
\frac{1}{N}\sum_{i=0}^{N-1}r_i
\]
 and the variance
\begin{equation} 
\sigma_N^2 = \frac{1}{N-1}\sum_{i=0}^{N-1}(r_i-\mu_N)^2
\label{eq:variance}
\end{equation}
If we sample, $M_s<N$, we can compute the average and variance and we
can compute $\sigma_M$. If we sample $M_s$ randomly, then
$\frac{\sigma_M}{\sigma_N} \sim \sqrt{\frac{M}{N}}$: in particular the
ratio $\frac{\sigma_M}{\sigma_N}$ should be distributed in the
vicinity of $\sqrt{\frac{M}{N}}$ as a normal standard distribution. We
can express the confidence that the sampling is in accordance with the
corpora by computing the probability
\begin{equation}
  2\Big(1 - \Phi(\frac{\sigma_M}{\sigma_N}-\sqrt{\frac{M}{N}})\Big).
\label{eq:sample}
\end{equation}
That is, the probability to be close to the expected variance, the
higher the better.

\mypar{Variance decrease} In principle, matching should decrease the
variance $\sigma_M < \sigma_N$; because we reduce $M<N$ and because we
remove outliers; if they do appear in both exposed and control, thus
they will not be outliers. We could choose the matching that reduces
the variance the most.  Unfortunately, the classic way to perform
matching will sample either exposed or control, not both; thus if the
variance is computed as in Equation \ref{eq:variance}, then this
criterion does not apply.

\mypar{ATT Response as normal distribution} One-to-one matching takes
one exposed $r^i_e$ and finds a control $r^j_c$ .  We can see that
\begin{equation}
E[r|E] - E[r|C] \sim \frac{1}{K}\sum_{i=0}^{K-1}r^i_e - r^i_c
\end{equation}
Thus we can create a distribution of the ATT response by $r^i_e -
r^i_c$. In general, the computation $r^i_e - r^i_c$ is actually $r^i_e
- w_ir^i_c$ where $w_i$ is a weight associated with the multiplicity
of the matched control and distance from exposed. For us, $w_i =1$
represents perfect match.  As such, we can compute average $\mu_E$ and
variance $\sigma_E$. If the distribution above is normal we can use
the ratio of average and variance to describe how well the matching
represents the final result: $2(1 - \Phi(|\mu_E|)/\sigma_E)$. 

In our cases, while the two moments specify completely an ideal
normal distribution, they do not give justice to the empirical
distribution. In fact, the normal distribution is {\em fatter} close
to the average and {\em thinner} at the tails.

\mypar{ATT Response as Laplace distribution} The Laplace distribution
provides a different approximation, sharper close to the average and
still symmetric.

\Doublefigure{0.5}{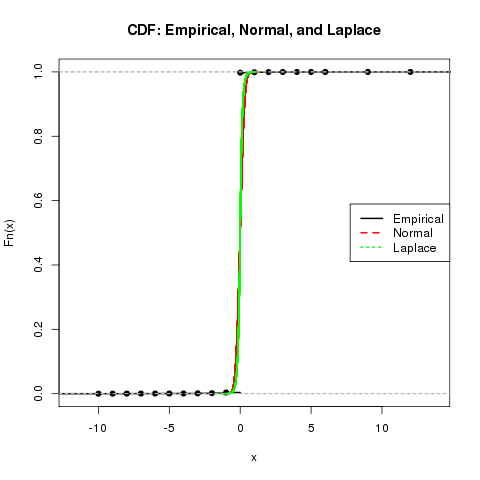}{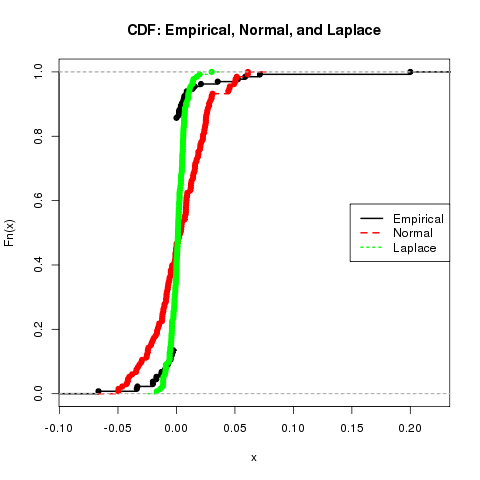}{Each
  element and Cluster contribution of the response
  distribution}{fig:distributions}

If we build a cluster, each cluster has its own response: we can
estimate the distribution of the response by the contribution of each
pair or by cluster. While using cluster responses, the distribution
may not be approximated by a Normal distribution nor by Laplace, see
Figure \ref{fig:distributions}.

\mypar{Skewness} Normal and Laplace distributions are poor
approximations of the tails of the empirical distributions. The
computation of the skewness describes with a single measure whether
these assumptions are appropriate and also whether or not we can trust
the confidence levels that we measure using the Normal or Laplace
distribution assumption. By construction, if the exposure is
effective, the distribution will be skewed and thus the normality
assumption will fail.

\mypar{$\chi^2_k$ distributions of users' response} Each cluster has a
collection of users and thus responses. The responses are correlated
in the clusters, by construction (or assumption). They should be
independent otherwise, across clusters. If the cluster has $k$
components, we can take the variance of each and their addition should
be the realization of the $\chi^2_k$ with $k$ degrees of freedom. Then we can
use the difference and the known distribution of the $\chi^2_k$ to
compute a confidence level, see Figure \ref{fig:chisquare}.

\singlefigure{0.5}{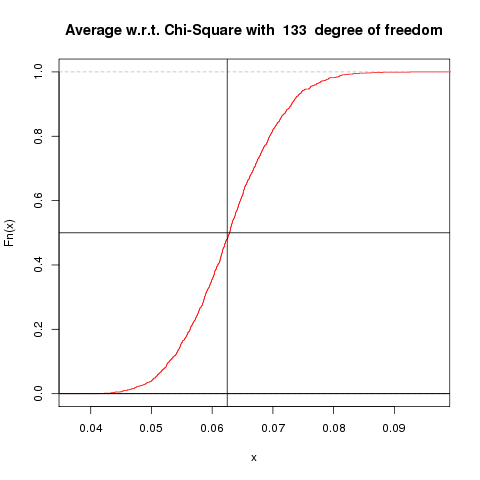}{Approximation by
  $\chi^2_k$}{fig:chisquare}

The main goal of these quality metrics and their description as
confidence level is to provide a measure of how well the clustering
and matching work. Basically, these metrics use two (or three) moments
of the response (average and variance) and the assurance of quality is
based on variations from an average, which comes from a known
distribution. Unfortunately, the response distribution seems to have a
mixed distribution and thus two moments cannot capture
this nature.

\mypar{Exposed and Control Entropy Constancy} The clustering/matching
should be independent of the response, just to avoid systematic bias.
One way to compare the quality of the matching is by measuring the
information contained in the original corpora and in the matched one.
Actually, we expect to have different distributions for exposed and
control.

We should and can compare the response distributions before and after
matching; in \cite{DAlbertoD2009}, we provide not only distance but
also confidence. Here, we simplify the problem and compute and compare
{\em entropy}. Take the exposed distinct response or the original
corpora
\begin{equation}
{\cal E}_e^N = -\sum_{\ell=0}P[r_e^\ell]\log_2(P[r_e^\ell])
\label{eq:entropy}
\end{equation}
${\cal E}_e^N$ is the expected information of the exposed
response. We can compute ${\cal E}_c^N$, which is the entropy of
control. We then can compare the entropy after the matching. Simply
put, we compute the pair:
\begin{equation}
({\cal E}_e^M{-}{\cal E}_e^N,{\cal E}_c^M{-}{\cal E}_c^N)
\label{eq:entropy2} 
\end{equation}
if the pair is positive, it means that the matching increased the
entropy and thus it censored responses with less
information. Otherwise, the matching is sampling responses with more
information. Among the matching algorithms, we should choose the one with the
smallest entropy difference for both exposed and control. In fact,
we are after the computation of Equation \ref{eq:2} where the
distribution affects the averages.

\mypar{Boostrap}  
We can consider the lift as a statistics and as such we can estimate a
variance using bootstrap. If the response $r(d,e)$ has $lift = E[r]$
and variance $\sigma^2_r = E[(r-lift)^2]$, then by bootstrapping the
lift as average $\hat{lift} = \frac{1}{N}\sum_i r_i$ has an average,
say $\mu_{lift}$, and $\sigma_\mu$.

Larger is the number of devices used in the experiment, smaller will
be the $\sigma_\mu$. This is because the average approximates the
expectation asymptotically 
\[\lim_{N\rightarrow \infty} (E[r]
- \frac{1}{N}\sum_i r_i )\rightarrow 0.\] As such, $\sigma_\mu$ and
$\sigma_r$ are not related: the response can have stable distribution
with unbounded variance $\sigma_r$ (i.e., have undefined variance for
$\alpha < 2$) and the $\sigma_\mu$ still will converge to zero because
the distribution has a finite expectation.

If we are prepared to run enough iterations, and considering that
$\sigma_\mu$ must be bound, we can use the equation $2(1 - \Phi(|lift
- \hat{lift}|)/\sigma_\mu)$ to achieve a p-value and thus a confidence
level.  Bootstrapping is also used to determine if the $lift$ of this
sample is a good approximation of the larger
population \cite{RePEc:idb:spdwps:1005}. For matching without
replacement, 1-1 matching, this is a welcome
approach \cite{Austin2014} and suggested to fill in the missing
information due to small experiments. For matching with replacement,
bootstrapping must be used carefully \cite{abadie:imbens:2008}. For
matching with replacement and large sample it may be not needed
because the distribution is known \cite{NBERw15301,AbadieImbens2006};
the application of bootstrap for continuous distribution of the
features and continuous response is an open question.

\mypar{Campaign as sample} Any matching algorithm samples the experiment to
remove the original bias, unfortunately the lift after matching does
not generalize to the original experiment because, in advertising,
targeting is not strongly ignorable and because the original
experiment is not a sample (it contains all exposed) and we are not
trying to estimate the lift for the hypothetical exposed.

\section{Experiments}
\label{sec:experiments}

In this section, we shall present our experimental results.  One way
to be able to use the methods above for large corpora is to sample
randomly. We show the hidden danger of this expedient
Section \ref{sec:sampling}. We then provide examples where all methods
provide information and thus none cannot be excluded a priori,
Section \ref{sec:comparison}. We show how the matching algorithms
presented here are comparable to the current available
Section \ref{sec:feature-algorithms}. We conclude in
Section \ref{sec:alg-performance} where we compare the average lift
computations for balanced approach of Section \ref{sec:CMS} and the
more general and unbalanced approach, showing how large experiments
can be matched.


\subsection{The danger of Sampling}
\label{sec:sampling} 
\begin{table}[hbt]
   \tbl{The different lifts by random sampling 200,000 users out of
      310,000\label{tb:sampling}}{%
      \small
      \begin{tabular}{|r|r|r|r|r|r|r|r|r|r|r|}
      
        \hline
        iteration&sort&sort-time&k-means&k-time&subclass& sub-time&exact&e-time& cem& c-time \\ 
        \hline \hline     
        1& -122 & -145 & -132 & -150 & -124  &  -88 & -100 & -132 & -100 & -132  \\
        2&  326 &  538 &  346 &  594 &  297 &  357 &  400 &  308 &  400 &  308 \\
        3& -101 & -673 &    0 &   26 & -163 &  166 &  185 & -209 &  185 & -209 \\
        4&   -3 &   14 &    3 &   -3 &   23 &   52 &   52 &   -3 &   52 &   -3 \\
        5&  -42 &  123 &  -10 &  -22 & -154 & -104 &  -70 &  -97 &  -70 &  -97 \\
        6& -118 &  -52 & -133 &  -59 & -156 & -109 &  -82 & -113 &  -82 & -113 \\
        7& -110 &  454 & -157 &  587 & -187 & -167 & -141 &  -43 & -141 &  -43 \\
        8&  -37 &  112 &  -47 &  166 & -134 & -113 & -111 &   38 & -111 &   38 \\
        9& 1637 &  941 & 1257 & 3795 &-1660 &-1019 &-1119 & -969 &-1119 & -969 \\
        10&  -63 &   -4 &  -79 &  -48 &  -84 &  -60 &  -88 &  -45 &  -88 &  -45 \\ \hline
        all &     -96 &  -91 &  -100 &  -50 &  -131 &  -71 &  -60 &  -88 &  -60 & -88 \\ \hline
        \hline
      \end{tabular}
    }
\end{table}

When we started our research, we applied known methods such as Match()
\cite{Sekhon2011}. To  cope with the long execution time for large
experiments, we sampled the users randomly but keeping the ratio. Our
experiments may have rare responses, fat tails; for features
represented in binary format, we have large classes creating
many-to-many matching; we noticed that sampling can bring forth
inconsistent results. For example, we took an experiment using a
balanced control-exposed set and discrete classes, the experiment had
about 310,000 users and we sampled it to 200,000. We ran 10 different
sampled matching and computed the lift ($(E[r|E]-E[r|C])/E[r|C]$) as
relative measure.  Also, we added features with time information
(i.e., two extra dimensions) to help the propensity score matching. We
summarize the results in Table \ref{tb:sampling}. Considering that in
Section \ref{sec:CMS} we suggest to sample Control, thus the response,
the table results should be a warning for large and small campaigns.

What is confusing with this experiment?  First, adding features does
not help provide consistent measures; what is lost during sampling is
lost and further space investigations seem helpless. Second, each run
provides quite different lifts in absolute values and in signs
(opposite).  In three out of ten iterations all matching results show
only negative lifts, one shows only positive lifts, and for six we
have mixed results. The experiment is not robust, but as we can see
the original experiment produce consistent, although negative,
results. In our scenario, the control group has much more signal as
the exposed group, thus sampling the experiment or sampling control
must be done with care and not randomly.

\subsection{Match comparisons}
\label{sec:comparison}

Let us take the example used in Section \ref{sec:sampling}. We do not
perform sampling and also we set to compute an unbalanced
matching. The number of users is exactly 302862. For each algorithm,
we use 25 and 27 features (whether or not we use time information). If
we use 25 features, the algorithm is fully specified by its name
(i.e., sort, k-means, subclass, exact and cem), if we use 27 features
we use longer names: sort-time/s-time, k-means-time/k-time,
subclass-time/s-time, exact-time/e-time and cem-time/c-time (features
are discrete again).

Match(), full and optimal from MatchIt() results are not reported
because they will take more than 3 hrs, which is not acceptable for
our purpose. The execution time to model the propensity score and to
score the user is about 8sec.  Also further tests show that their
performance increases linearly with the number of users in the
matching tests (considering the number of features fixed).

In Table \ref{tb:entropy}, we show the performance of the same tests,
 the entropy difference (see Equation
\ref{eq:entropy2}), and the standard deviation of the users'
responses. The corpora has variance of 0.0980. Notice that our methods
tend to decrease the entropy but in absolute value, this is the minimum.

\begin{table}[hbt]
    \tbl{Unbalanced computation: difference in entropy and variance\label{tb:entropy}}{
    \centering \small
    \begin{tabular}{|l|l|l|l|}
      \hline
      algorithm& exposed & control & variance \\ \hline \hline
      sort           & {\bf -0.00016} & {\bf -5.91886e-05} & 0.0978\\
      sort-time      & -0.00056     &-0.00061     & {\bf 0.0959}\\
      k-means        & -6.45606e-05 &-6.06477e-05 & 0.0979\\
      k-means-time   &  0.00281     &-0.00028     & 0.0970\\
      subclass       &  0           & 0.00404     & 0.0975 \\
      subclass-time  &  0           & 0.00600     & 0.0965\\
      exact          &  4.52999e-05 & 0.00808     & 0.1009\\
      exact-time     &  0.00252     & 0.01500     & 0.0978\\
      cem            &  4.52999e-05 & 0.00808     & 0.1009\\ 
      cem-time       &  0.00252     & 0.01500     & 0.0978\\ 
      \hline \hline     
    \end{tabular}
}
\end{table}

If we use the moments and a few assumptions about the users' responses,
we can compute the probability such that the empirical distribution
is indistinguishable from the assumed distribution, see Table
\ref{tb:confidence}; then, all matching algorithms accept the equality
assumption. Sort and k-mean use the clusters: by using the Normality
assumption the matching will be accepted, using the Laplace assumption
the matching will be rejected. Other methods use single user response
and thus due to the number of users (300,000 users) they converge to
normality and the others follow.
\begin{table}[hbt]
  \tbl{Unbalanced computation: confidence in the matching process\label{tb:confidence}}{
  \centering \small
  \begin{tabular}{|l|l|l|l|l|}
    \hline
    Algorithm    &Normal & Laplace & $\chi^2_k$ & Eq. \ref{eq:sample}  \\ \hline \hline  		  
    sort              &0.99911 &0.95689 &0.94941 &0.99943 \\
    sort-time         &0.99798 &0.72556 &0.99178 &0.98933 \\
    kmeans            &0.99824 &0.67636 &0.97187 &0.99940 \\
    kmeans-time       &0.99749 &0.65775 &0.98537 &0.98061 \\
    subclass          &0.99867 &1       &0.99825 &0.99581 \\
    subclass-time     &0.99927 &1       &0.99926 &0.98781 \\
    exact             &0.99941 &1       &0.99958 &0.97563 \\
    exact-time        &0.99913 &1       &0.99910 &0.97086 \\
    cem               &0.99941 &1       &0.99846 &0.97563 \\
    cem-time          &0.99913 &1       &0.99936 &0.97086 \\
    \hline \hline     
  \end{tabular}
}
\end{table}

\subsection{Audience Selection, Features, and Algorithms (Cor)Relation}
\label{sec:feature-algorithms}
\label{sec:features}

In our experience, our problem space has three basic dimensions: First
is the choice of the exposed group and thus control; Second is the
dimension number and quality of the feature space describing our
audience; Third is the set of algorithms and what they can expose for
all the above. Eventually, we would like to infer recommendations
about what works, especially at scale. In this section, the largest
campaign has ten million devices and the smallest a few hundred
thousands.

In this section, we consider a few dozen campaigns and we applied a
balanced approach, Section \ref{sec:CMS}, with discrete feature space
Section \ref{sec:user-profile} using only registration data, known as
prior. In this scenario, the user response is discrete covering a set
of discrete values. The feature space specifies a discrete space,
although possibly large, it is limited and users could be clustered
into a few thousands classes. The estimate of the targeting function
reflects the discrete space nature; thus, the matches are often
many-to-many and the exposed group has priority, that is, we sample
control. For all the matching algorithms, there is always a
many-to-many matching because the calipers will infer classes and
within any class we do not apply a {\em nearest} matching. Even for
the standard algorithms, they apply a 1-to-many matching introducing
weights.

\doublefigure{0.80}{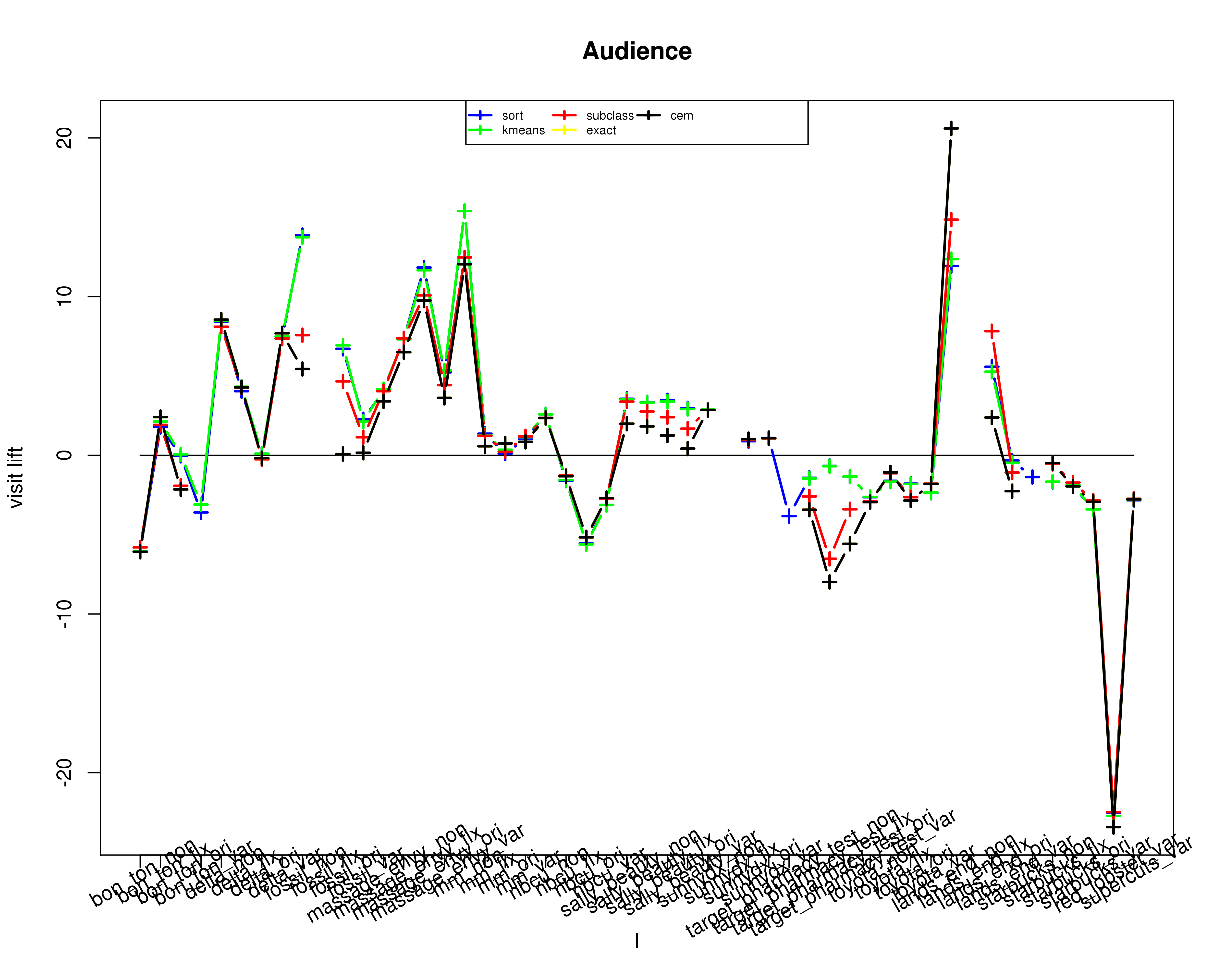}{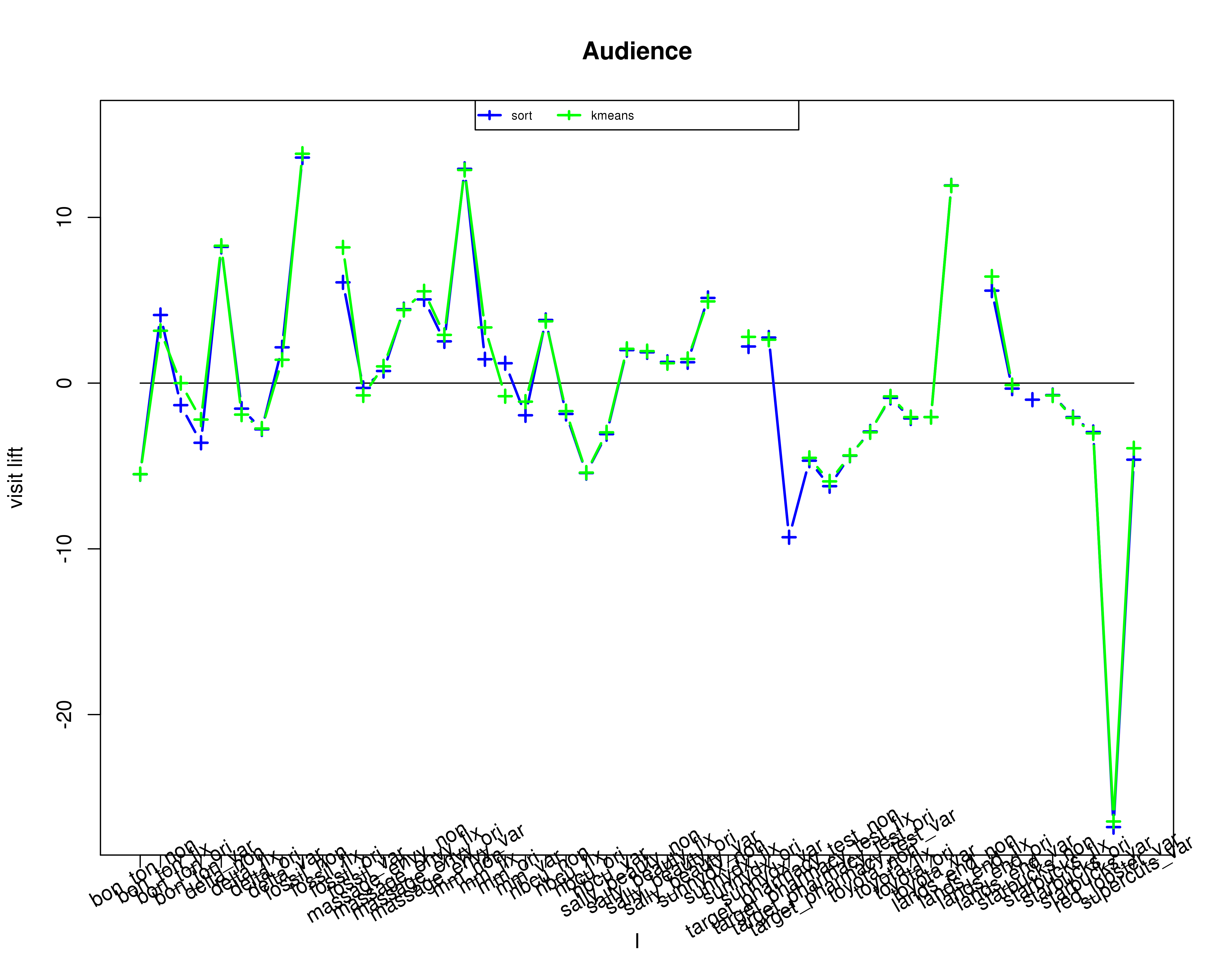}{Audience keywords: Unbalanced $k$-to-$m$ matching (above) and Balanced many-to-many matching (below))}{fig:keywords2}

In Figure \ref{fig:keywords2}, we compare our algorithms together with
more standard ones. We can appreciate that the final lift values
differ little. This simple experiment shows that our matching
algorithms are equivalent to others with the advantage that can be
applied to larger campaigns without loss of accuracy. We make sure
that the matching as we designed and developed does not loose
information for the type of our experiments. In practice, we show
experimentally that our matching are a useful contribution to the
literature especially as meaningful extension, if not the only
available extension. In the following section, we go even bigger. To
do so, we need a different framework and we need to use an unbalanced
approach.

\subsection{Experiments and analysis of stochastic hits} 
\label{sec:brownian}                       

Setting a specific radius or the contour of a parcel as a boundary for
the computation of a hit is simple to explain and to use. However, how
we can account for those impressions close by those boundaries by
users we do not see inside those same boundaries. They may have
stepped out of the location we are interested in and sent us a
signal. In this section, we discuss the application of what we called
stochastic hit, if an impression is close enough to a location we may
consider to give it a probability (of a hit) by using a known
distribution such as the IG presented in Equation \ref{eq:ig} or a
{\em lognormal} distribution.

We consider two campaigns A and B, they have more than thirty
locations of interests relatively sparse geographically. One campaign
is to advertise a car company and the other is a restaurant chain. 

We create these two experiments for both campaigns: we consider $R=30$
and $R=96$ meters, $R$ is the radius we use to consider an impression
a hit. For each experiment, we consider all the users that have a hit
at distance $d<=R$. Then we count the number of times and the
distances $0\leq d \leq 10R$ with a precision of one meter. 

\Doublefigure{0.5}{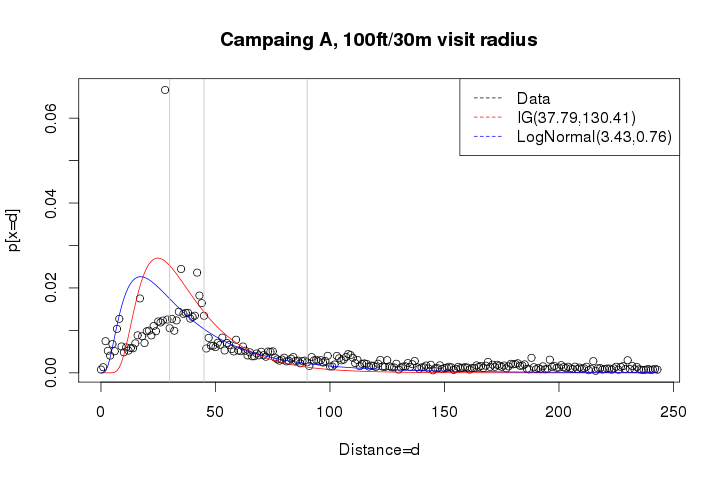}{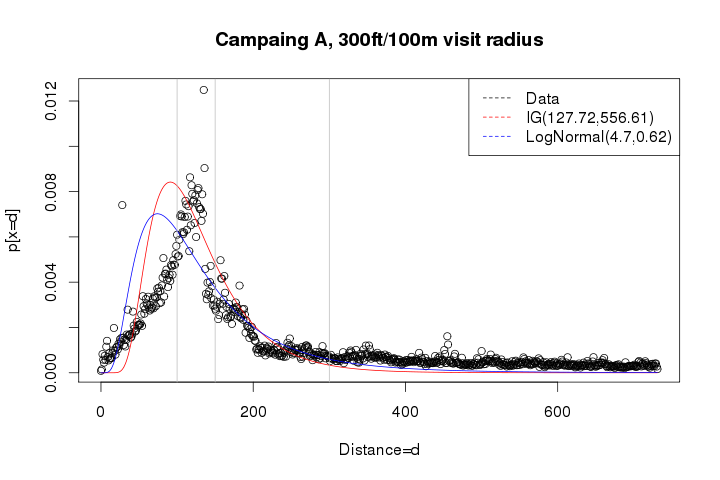}{Campaign A: Conditional distance distribution with users who has a visit within 30 and 100 meters}{fig:global21015}

In Figure \ref{fig:global21015}, we present the two experiments when
we observe the distribution of who hit ($d<R$ for $R=30$ and $R=100$
meter). Within $0\leq d \leq R$, we have investigated a simple
polynomial curve fit and we have found that $p[x=d]\sim \alpha
d$. This simple observation means that $P[x\leq R] \sim R^2$ and,
thus, the number of hits is proportional to the area. This is not true
for each location, but for the aggregate of 30 or more locations, we
achieve such a nice property, which is intuitive.

\Doublefigure{0.5}{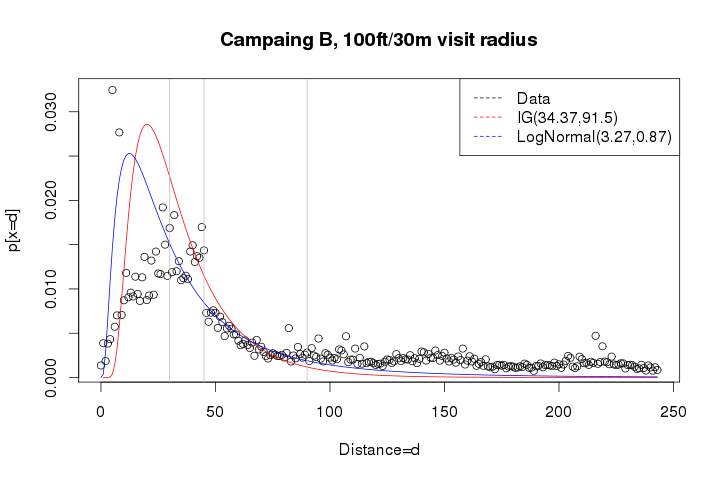}{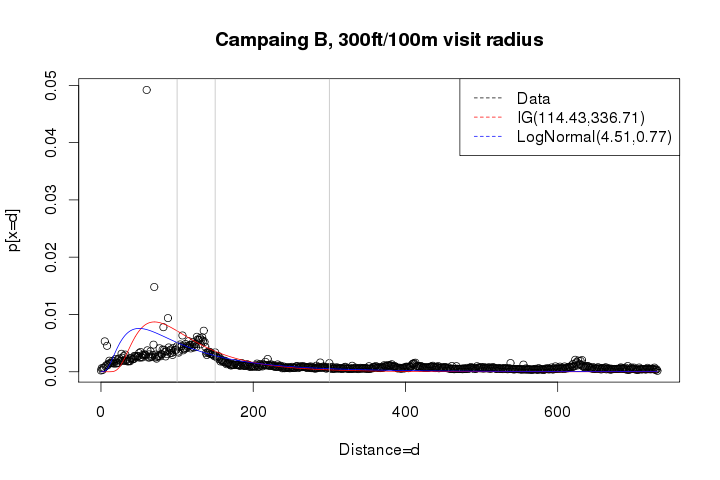}{Campaign B: Conditional distance distribution with users who has a visit within 30 and 100 meters}{fig:global20663}

In Figure \ref{fig:global20663}, we present the second campaign and we
can appreciate that the experiment have similar results. The slope is
different, specific to the campaign and locations set,
$p[x=d]\sim \beta d$. In principle, we can fit multiple distribution
models: we fit the inverse Gaussian (IG) and a log-normal. We plot the
correspondent distribution for the models. 

Our goal is to estimate the probability that an impression with
distance $d>R$ could be a probability of a hit. In practice, we could
use $IG(\mu, \lambda)$ to estimate the probability for every
impression in the range $[R, R+\frac{R}{2}]$, in
Figure \ref{fig:global21015} and \ref{fig:global20663} we represent
this space by the first two vertical gray lines (from the far
left). And we could use the $LogNormal(\mu,\sigma)$ for the interval
$(R+\frac{R}{2}, 3R]$.

\subsection{Large scale  lift comparison}
\label{sec:alg-performance} 

To the best of our knowledge, the system we present in
Section \ref{sec:CMS} and show results in
Section \ref{sec:feature-algorithms} is the only one capable to tackle
an experiment with 2 million users in any practical way.  In this
section, we present a different prospective in order to create and to
measure even larger experiments and show how sampling is still a
lingering issue, although for different reasons.

Let us introduce the {\bf impression space}: We formally introduced
the concept of impression and we used to specify visits as in
Equation \ref{eq:visit}. Where do these impressions come from? We are
listening to a fire-hose of streaming impressions that we can bid
through a collection of exchanges. Our budgets dictates how much we
can listen and it changes. Here, we call this fire-hose real time bid
(RTB) exchanges, the volume of RTB is a function of an allocation
budget and it will change as a function of company wide budget,
hardware allocation and hardware/software failure. The RTB is composed
of three non intersecting parts:
\begin{itemize}

\item Listening RTB (LRTB) is a random sample of the fire-hose used only for collection purpose, let us say that LRTB is 10\% of the RTB. 

\item Won RTB (ND),  we bid and we win the impression, thus we deliver our advert. The volume of impressions here is a function of campaign budget and pacing. 

\item Unwanted RTB (URTB) is the remaining impressions, the larger portion of RTB.   

\end{itemize}

Using all URTB, LRTB, and ND (i.e., urtb-lrtb-nd) impressions and our
unbalanced approach, we will not sample impressions, visits, nor
users. This is the ultimate representation of our experiment space. As
such, it puts quite a few practical and economical constraints in the
experiment measure. For example, a national campaign like Starbucks
counting 10 thousands locations and three months period, will touch
approximately 100 million devices and (hundred of) billions
impressions. This experiments will have maximum number of users and
visits.

Historically and thus in Section \ref{sec:CMS}, the balanced approach
uses the URTB and ND impressions, it intersects the exposed users to
those impressions and it samples the control, (i.e., c-urtb-nd).  For
national campaigns as above, the sheer size of impressions to manage
can be quite large. As a practical effect, we sample control, we
sample visitors, and thus we sample visits.

Assume we embrace a different sampling: sampling of impression in
time. The LRTB has the property of being an unbiased and random sample
of RTB, thus the space LRTB-ND (i.e., lrtb-nd) may have all the
information we need, a critical size to compare users and visits, and
a practical size to have the experiment measures in a more economical
way.

A long tail distribution is applicable here. There are a lot of users
with few impressions, there are a lot of users with zero visits. The
c-urtb-nd samples mostly control, even though we use the terminology of
balanced approach, we can appreciate the irony of the name, if or when
we are control. The lrtb-nd approach will sample the impressions, we
will pick users with enough impressions, but we do that without any
bias to the targeting. Of course, a critical mass has to be met by
both, otherwise the sampling curse will be visible at this level as
well. Now, we can introduce the final experiment.

We took twelve campaigns of various sizes. For each we measure lift
without any matching, we use the term of general lift. These
experiments have different goals. In Figure \ref{fig:general-1}, we
show the results. 

\singlefigure{0.90}{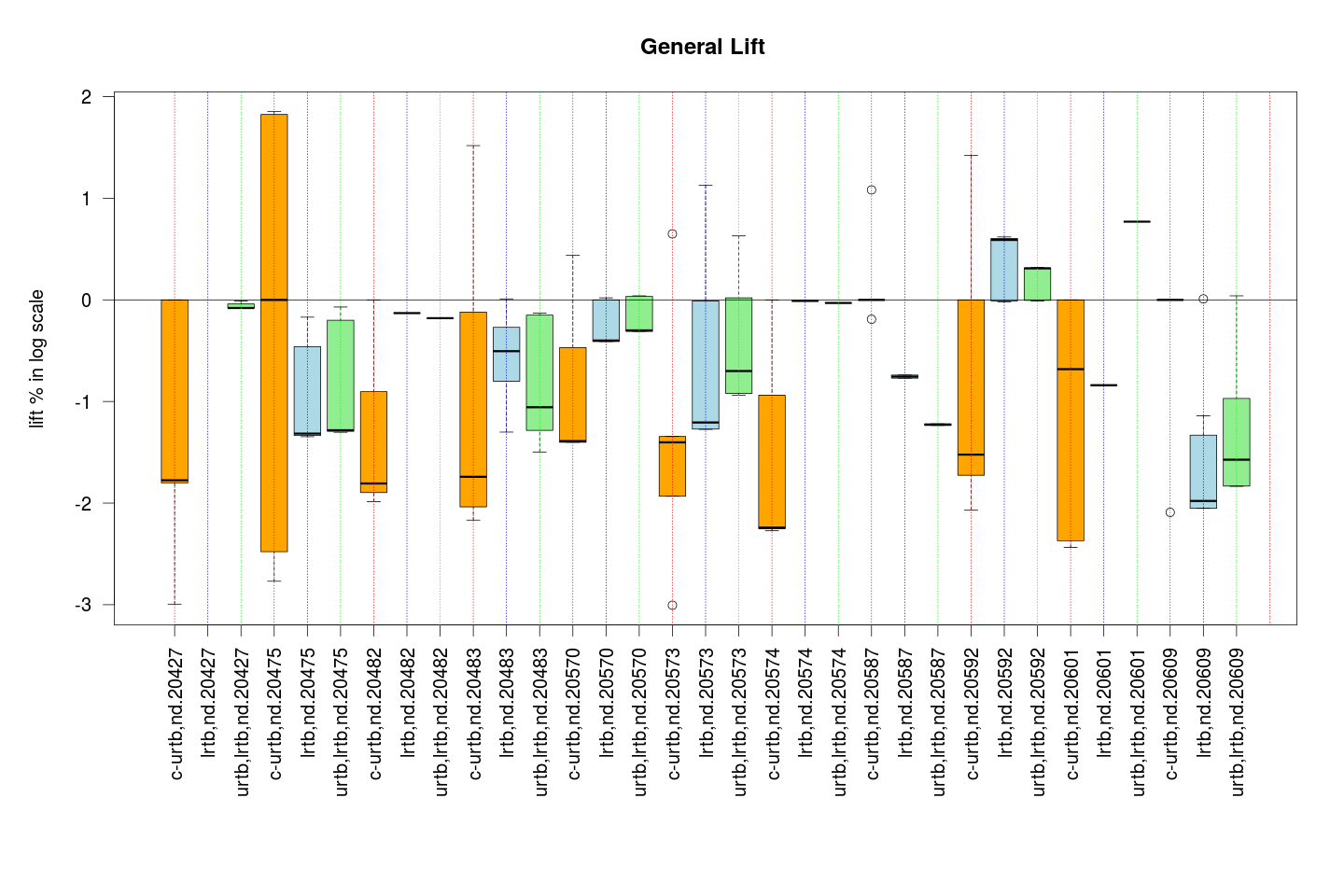}{General Lift using different methodology and sampling}{fig:general-1} 
To provide a clear presentation of the small differences, we opted for
a non standard box plot. For any lift in the range $[-1,1]$, we
present it as it is. For any lift in the range $[1,\infty)$, we
represent it as $1+\log_{10}(lift)$. We work out similarly the
negative lift smaller than $-1$. Thus, we use a logarithmic
scale but only for large enough lifts.

We can appreciate that the general lift for all campaigns is negative
(on an average). There is only one exceptions where c-urtb-nd is not
correlated to urtb-lrtb-nd (i.e., campaign $20592$), and only one
exception for rtb-nd (i.e., campaign $20601$). This means that
sampling can be an issues even at this scale but it is moderate. Why
is it always negative? The main reason is targeting: targeting is far
from being ignorable and the control group is not really
comparable. We need to apply matching and we show the lift comparison
in Figure \ref{fig:general-2}.

\singlefigure{0.90}{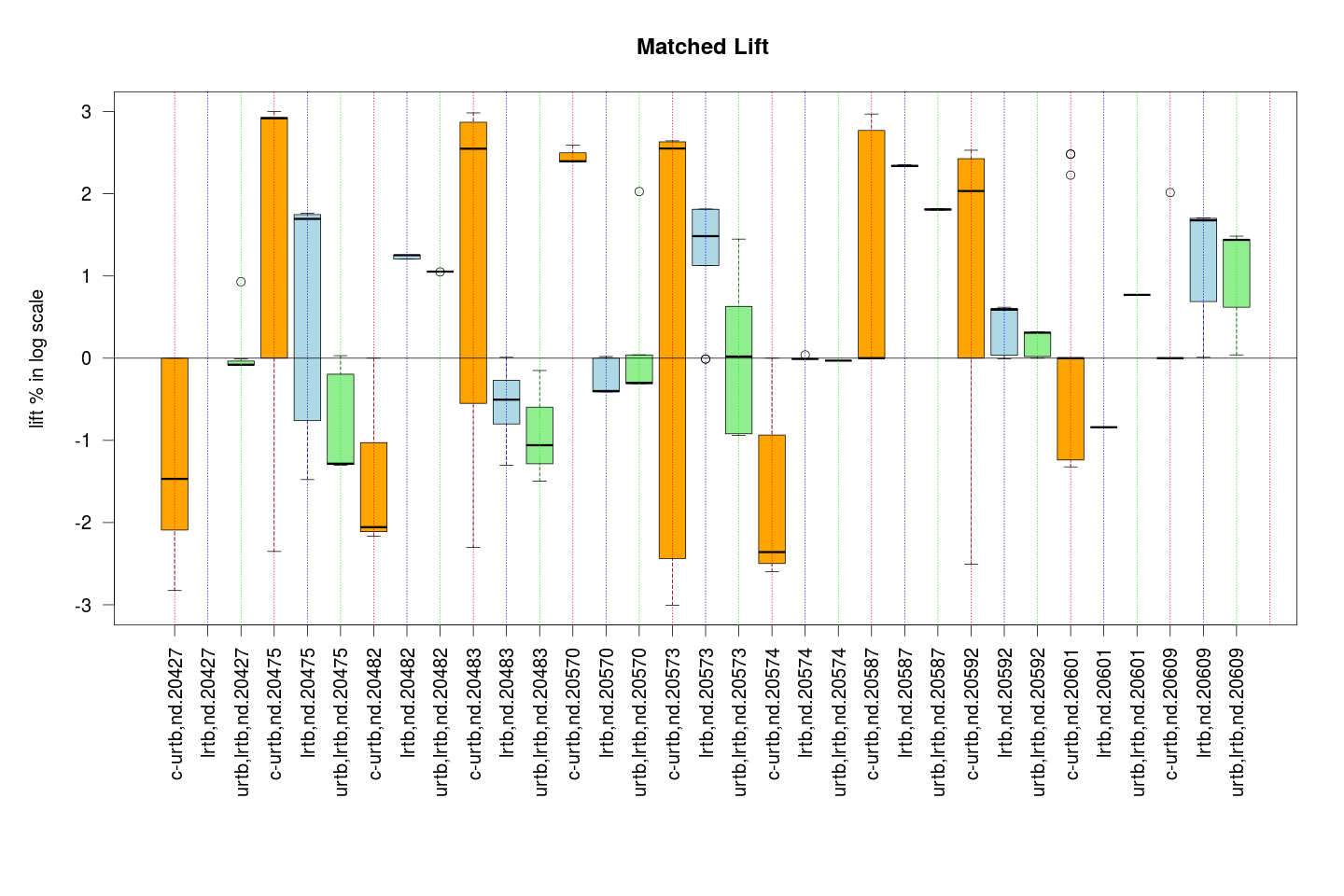}{Effect of Matching}{fig:general-2}

There is even a stronger correlation on the sign of the lift while
applying matching: only campaign $20483$ is the exception. In general,
sampling increases the variance of the lifts (a little obscured by the
log scale) and the absolute value of the lift; that is, sampling
reduce the population of the experiments, making the visit per user
effect larger. 

The larger the experiment is, then the smaller the lift is.  This is
an important and practical consideration. Lift is basically a
comparison measure with respect to a control that it can be much
larger than the exposed group and thus the control visits can be much
larger. Eventually, we are bound to measure lift as number of relative
visits with respect to all visits, if all visits increases the lift
will decrease. This is the course of relative measures: A larger
campaign may have a larger effect (more visits) than a smaller one,
but relatively to the population it reached, the larger campaign will
have smaller lift.

\section{Conclusions}
\label{sec:conclusions}
In this work, we present a common and important problem for
advertising companies: how to quantify the effect of a digital
campaign. Specific to our field, we need to measure visits and the
speed of visits for an exposed group with respect to a control
baseline. We introduce a general approach and we explain two different
methodologies using a balanced and an unbalanced exposed-control
selection. We follow through by presenting two implementations and
showing their different capabilities and similarities. We show that we
can write scalable matching algorithms that can be practical, accurate
as much as the ones available in the literature. We show that our
algorithms can be applied to very large experiments.

After all, the two methods should agree on an average
especially if the experiments are well deployed. Our goal is to share
our intuitions, our development solutions, and insights. This is a
complex problem: our solutions have often been driven by practical
necessities, limited resources, and clear goals.  Here we show our
best and always moving effort to present our understanding and shed
some light to possible, sound, and practical solutions.

\bibliographystyle{acmsmall}

 \bibliography{pa.strass2}

\end{document}